\documentclass[11pt,a4paper]{article}
\usepackage{jheppub}

\usepackage{amsmath}
\usepackage{amssymb}
\usepackage{graphicx}
\usepackage{rotating}
\usepackage{color}
\usepackage{multirow}
 \usepackage[T1]{fontenc}

\def\bY{{\bf Y}}
\def\bA{{\bf A}}

\def\bU{{\bf U}}

\def\bmm{{\bf m}}
\def\bM{{\bf M}}
\def\bR{{\bf R}}
\def\hatbmm{{\bf \bf \hat{m}}}
\def\hatbM{{\bf \bf \hat{M}}}
\def\bL{{\bf L}}
\def\GeV{{\rm GeV}}
\def\TeV{{\rm TeV}}

\newcommand{\nn}{\nonumber}

\def\gsim{\raise0.3ex\hbox{$\;>$\kern-0.75em\raise-1.1ex\hbox{$\sim\;$}}}
\def\lsim{\raise0.3ex\hbox{$\;<$\kern-0.75em\raise-1.1ex\hbox{$\sim\;$}}}

\title{Constrained SUSY seesaws with a 125~GeV Higgs}

\author[a]{M.~Hirsch}
\author[b]{F. R. Joaquim}
\author[c]{A. Vicente}

\affiliation[a]{AHEP Group, Instituto de F\'{\i}sica Corpuscular -- C.S.I.C./Universitat de Val{\`e}ncia, Edificio de Institutos de Paterna, Apartado 22085,
E--46071 Val{\`e}ncia, Spain}

\affiliation[b]{Departamento de F\'{\i}sica and Centro de F\'{\i}sica Te\'{o}rica de Part\'{\i}culas, Instituto Superior T\'{e}cnico, Universidade T\'ecnica de Lisboa, Av. Rovisco Pais, 1049-001 Lisboa, Portugal}

\affiliation[c]{Laboratoire de Physique Th\'eorique, CNRS -- UMR 8627, Universit\'e de Paris-Sud 11, F-91405 Orsay Cedex, France}

\emailAdd{mahirsch@ific.uv.es}
\emailAdd{filipe.joaquim@ist.utl.pt}
\emailAdd{avelino.vicente@th.u-psud.fr}


\abstract{Motivated by the ATLAS and CMS discovery of a Higgs-like
boson with a mass around 125 GeV, and by the need of explaining
neutrino masses, we analyse the three canonical SUSY versions of the
seesaw mechanism (type I, II and III) with CMSSM boundary
conditions. In type II and III cases, SUSY particles are lighter than
in the CMSSM (or the constrained type I seesaw), for the same set of
input parameters at the universality scale. Thus, to explain $m_{h^0}
\simeq 125~\GeV$ at low energies, one is forced into regions of
parameter space with very large values of $m_0$, $M_{1/2}$ or
$A_0$. We compare the squark and gluino masses allowed by the ATLAS
and CMS ranges for $m_{h^0}$ (extracted from the 2011-2012 data), and
discuss the possibility of distinguishing seesaw models in view of
future results on SUSY searches. In particular, we briefly comment on
the discovery potential of LHC upgrades, for squark/gluino mass ranges
required by present Higgs mass constraints. A discrimination 
between different seesaw models cannot rely on the Higgs mass data 
alone, therefore we also take into account
the MEG upper limit on ${\rm BR}(\mu \rightarrow e \gamma)$ and show
that, in some cases, this may help to restrict the SUSY parameter
space, as well as to set complementary limits on the seesaw scale.}

\begin{document}
\maketitle

\section{Introduction}
\label{sec:intro}

With the data accumulated in 2011 and 2012, both the CERN ATLAS and
CMS collaborations have recently claimed the discovery of a new
particle that resembles very much the long-awaited Higgs boson. The
mass of this new state, measured in good accordance in different decay
channels, is in the ballpark of $m_{h^0} \simeq (123-127)~\GeV$. While
the overall significance in the 2011 data was only $2.2\sigma$ in
ATLAS~\cite{ATLAS:2012ae} and $2.1\sigma$ in
CMS~\cite{Chatrchyan:2012tx}, with the 2012 update both experiments
increased their statistical significances to the $5\sigma$ discovery
threshold~\cite{CMStalk,ATLAStalk}. Especially noteworthy is that both
ATLAS and CMS observe an excess of events in the $\gamma\gamma$ and
$ZZ$ decay channels with an invariant mass which differs by roughly 2
GeV, i.e. consistent at the $1\sigma$ level. Complementary evidence
has been reported by the CDF and D0 experiments at the Tevatron. These
collaborations have recently released updated combined results on
searches for the Higgs boson~\cite{Tevatron}, finding a $\sim 3\sigma$
statistical significance in the $b \bar{b}$ decay channel.

Given that supersymmetry (SUSY) has been the most popular paradigm for
physics beyond the standard model (SM) in the last decades, the recent
LHC results have triggered the expected flurry of theoretical activity
dedicated to the study of how a relatively heavy Higgs constrains the
supersymmetric parameter
space~\cite{Hall:2011aa,Baer:2011ab,Feng:2011aa,Heinemeyer:2011aa,Arbey:2011ab,Arbey:2011aa,Draper:2011aa,Moroi:2011aa,Carena:2011aa,Ellwanger:2011aa,Buchmueller:2011ab,Akula:2011aa,Kadastik:2011aa,Cao:2011sn,Arvanitaki:2011ck,Gozdz:2012xx,Gunion:2012zd,Ross:2011xv,FileviezPerez:2012iw,Karagiannakis:2012vk,King:2012is,Kang:2012tn,Chang:2012gp,Aparicio:2012iw,Roszkowski:2012uf,Ellis:2012aa,Baer:2012uya,Baer:2012uy,Desai:2012qy,Cao:2012fz,Maiani:2012ij,Cheng:2012np,Christensen:2012ei,Vasquez:2012hn,Ellwanger:2012ke,Gogoladze:2012ii,Ajaib:2012vc,Brummer:2012ns,Ross:2012nr}. The
general consensus is that a lightest Higgs boson with a mass of
$m_{h^0} \sim 125~\GeV$ is uncomfortably heavy for {\em minimal}
SUSY. Here, by minimal SUSY we mean a supersymmetric model with no new
superfields and no new interactions, gauged or non-renormalizable, at
the electroweak scale. In this framework, the hefty Higgs requires either
multi-TeV scalar tops or very large stop mixing
\cite{Arbey:2011ab,Baer:2011ab,Baer:2012uya,Buchmueller:2011ab,Ellis:2012aa}. In
the latter case, the lightest stop could still be relatively light,
say $m_{{\tilde t}_1} \gsim 500~\GeV$~\cite{Benbrik:2012rm}. For a
constrained minimal supersymmetric standard model (CMSSM) with
universal boundary conditions at a high scale, such a spectrum
requires that at least one of the three basic parameters $M_{1/2}$,
$m_0$ or $A_0$ takes a minimum value of several
TeV~\cite{Arbey:2011ab,Baer:2011ab,Ellis:2012aa}. In addition, it has
been found that a moderately large $\tan \beta$ may be helpful to
increase the Higgs boson
mass~\cite{Arbey:2011ab,Baer:2011ab,Ellis:2012aa}.

The naturalness problem of the MSSM with a $\sim 125~\GeV$ Higgs mass
has revived the discussion around non-minimal supersymmetric
extensions of the standard model. In particular, the recent LHC data
has been scrutinized in the context of SUSY models with new F-terms
(like the NMSSM)~\cite{Ellwanger:2011aa,King:2012is,Gunion:2012zd,
Cao:2012fz,Vasquez:2012hn}, extended gauge models with additional new
D-terms~\cite{Batra:2003nj,Maloney:2004rc,Hirsch:2011hg,
Hirsch:2012kv,An:2012vp,Randall:2012dm},
heavy-SUSY scenarios like Split
SUSY~\cite{ArkaniHamed:2004fb,Dimopoulos:1995mi,Alves:2011ug},
``natural
SUSY''~\cite{Dine:1990jd,Baer:2012up,Papucci:2011wy,Craig:2012di,Baer:2012uy}) and
high-scale SUSY~\cite{Hall:2009nd,Giudice:2011cg}) or ``effective''
SUSY, i.e. SUSY with new non-renormalisable
operators~\cite{Polonsky:2000rs,Brignole:2003cm,Casas:2003jx,Dine:2007xi},
among others. In this work, we will follow an alternative approach and
assume SUSY is realized minimally. We explore the consequences of the
LHC Higgs search data on the CMSSM parameter space and the SUSY
spectrum, from a viewpoint similar to that taken in MSSM-dedicated
studies like, for instance, the one of
Ref.~\cite{Arbey:2011ab}. However, our analysis differs from these by
considering that a seesaw mechanism for neutrino mass generation is
implemented in the MSSM. Our motivation lies in the fact R-parity
conserving MSSM (with or without CMSSM boundary conditions) does not
provide an explanation for the observed neutrinos masses and, thus,
is not complete.

From the theoretical point of view, implementing the seesaw mechanism
in the (supersymmetric) SM seems to be the simplest (and most
motivated) solution to the neutrino mass problem. With renormalizable
interactions only, there are three tree-level realizations of the
seesaw mechanism~\cite{Ma:1998dn} usually called typeI~\cite{Minkowski:1977sc,seesaw,Mohapatra:1979ia,Schechter:1980gr,Schechter:1981cv},
II~\cite{Konetschny:1977bn,Schechter:1980gr,Marshak:1980yc,Lazarides:1980nt,Mohapatra:1980yp,
Cheng:1980qt,Schechter:1981cv}
and III~\cite{Foot:1988aq}. These three variations differ from each
other by the nature of their seesaw messengers. Namely, in type I an
effective neutrino mass operator arises from the decoupling of heavy
neutrino singlets, while in type II one integrates out a heavy SU(2)
scalar triplet with hypercharge two. Instead, in the type III seesaw
neutrino masses are generated through the tree-level exchange of SU(2)
fermionic triplets of zero hypercharge. If in type II and III one
extends the MSSM by just adding the superfields required to generate
neutrino masses, then one of the most appealing properties of the MSSM is
lost: gauge coupling unification. This stems from the fact that both
the scalar and fermionic triplets belong to incomplete SU(5)
representations. Unification can be easily restored by embedding those
states in full SU(5) multiplets like 15-plets in the case of type
II~\cite{Rossi:2002zb} or 24-plets~\cite{Buckley:2006nv} in the case
of type III. Note that, in addition to the SU(2) triplet, the 24 of
SU(5) contains a singlet which also contributes to the effective
neutrino mass operator and, thus, the decoupling of the 24-plet leads
to an admixture of type I and type III seesaws.

The main purpose of this work is to investigate whether imposing a
Higgs mass around 125 GeV allows to some extent to differentiate the
CMSSM from the constrained SUSY seesaws and also whether type II and III
seesaws are distinguishable among themselves. We will complement this
analysis by imposing the MEG constraint on the branching ratio of the
radiative lepton flavour violating decay $\text{Br}(\mu\to e\gamma)\le
2.4\times 10^{-12}$~\cite{Adam:2011ch}.

The rest of this paper is organized as follows. We start by recalling
the general features of the aforementioned SUSY seesaw models in
Section~\ref{sec:models} and present some discussion related with lepton
flavour violation (LFV) in Section~\ref{sec:LFV}. Afterwards, we
describe our numerical analysis and present its results in
Sections~\ref{subsec:numsetup} and \ref{sec:results}, respectively. Our
conclusions are drawn in Section~\ref{sec:conc}.

\section{Models}
\label{sec:models}

In the following we will briefly describe the three types of SUSY seesaw mechanisms considered in this work and possible embedding in a grand-unified (GUT) model based on the SU(5) gauge group. We use standard notation for the MSSM superfields, namely $L$, $Q$ and $H_u$ ($H_d$) denote the lepton, quark and hypercharge one (minus one) Higgs superfields, while the lepton and quark singlets are $E^c$, $D^c$ and $U^c$. The vacuum expectation values of $H_{u,d}$ are denoted by $v_{u,d}/\sqrt{2}$ with $\tan\beta=v_u/v_d$ and $v=\sqrt{v_u^2+v_d^2}=246~\GeV$.

\subsection{Supersymmetric seesaw type I}
\label{sec:modelI}

In the case of the supersymmetric type I seesaw, very heavy singlet superfields $N^c$ are added to the MSSM, yielding the following superpotential below the grand-unification scale $M_{GUT}$:
\begin{eqnarray}
\label{eq:WtypeI}
W_{I} &=& W_{MSSM} + W_{\nu} \label{eq:superpotI} \thickspace, \\
W_{MSSM}& = & \bY_u U^c Q H_u
         - \bY_d D^c Q H_d
         - \bY_e E^c L H_d
              + \mu H_u H_d \thickspace, \\
 W_{\nu}& = & \bY_\nu N^c L H_u
          + \frac{1}{2} \bM_R N^c N^c \thickspace,
\end{eqnarray}
where SU(2)-invariant products are implicit. This model can be realized in an SU(5) framework taking the following SU(5) matter representations: $ 1= N^c$, ${\bar 5}_M=\{D^c,L\}$
and $10_M=\{Q,U^c,E^c\}$.
At the effective level, a dimension five neutrino mass operator of the type $LLH_uH_u$ originates from the decoupling of the heavy singlets, leading to an effective neutrino mass matrix given by the well-known seesaw formula
\begin{equation}
\bmm_\nu = - \frac{v^2_u}{2} \bY^T_\nu \bM^{-1}_R \bY_\nu\,,
\label{eq:mnuI}
\end{equation}
after electroweak symmetry breaking (EWSB).
Being complex symmetric, $\bmm_\nu$ is diagonalized by a $3\times 3$ unitary matrix $\bU$~\cite{Schechter:1980gr}
\begin{equation}\label{diagmeff}
\hatbmm_{\nu} = \bU^T  \bmm_{\nu} \,\bU\,.
\end{equation}
The lepton mixing matrix $\bU$ can be parameterized in the standard form
\begin{eqnarray}\label{def:unu}
\bU=
\left(
\begin{array}{ccc}
 c_{12}c_{13} & s_{12}c_{13}  & s_{13}e^{-i\delta}  \\
-s_{12}c_{23}-c_{12}s_{23}s_{13}e^{i\delta}  &
c_{12}c_{23}-s_{12}s_{23}s_{13}e^{i\delta}  & s_{23}c_{13}  \\
s_{12}s_{23}-c_{12}c_{23}s_{13}e^{i\delta}  &
-c_{12}s_{23}-s_{12}c_{23}s_{13}e^{i\delta}  & c_{23}c_{13}
\end{array}
\right)
 \left(
 \begin{array}{ccc}
 e^{i\alpha_1/2} & 0 & 0 \\
 0 & e^{i\alpha_2/2}  & 0 \\
 0 & 0 & 1
 \end{array}
 \right)\,,
\end{eqnarray}
with $c_{ij} = \cos \theta_{ij}$ and $s_{ij} = \sin \theta_{ij}$. The
angles $\theta_{12}$, $\theta_{13}$ and $\theta_{23}$ are the solar, the reactor (or CHOOZ) and the atmospheric neutrino mixing angle, respectively, while $\delta$ is the Dirac phase and
$\alpha_{1,2}$ are Majorana phases.

It is well known that the Dirac neutrino Yukawa couplings $\bY_\nu$
can be defined in terms of the physical neutrino parameters, up to an
orthogonal complex matrix $\bR$~\cite{Casas:2001sr},
\begin{equation}\label{Ynu}
\bY_{\nu} =\sqrt{2}\frac{i}{v_u}\sqrt{\hatbM_R} \bR \sqrt{{\hatbmm_{\nu}}} \bU^{\dagger},
\end{equation}
where $\hatbmm_{\nu}$ and $\hatbM_R$ are diagonal matrices containing the light and heavy neutrino masses, respectively. It is worth noting that, in the special case of $\bR={\bf 1}$, the non-trivial flavour structure of $\bY_{\nu}$ stems from the lepton mixing matrix $\bU$.

\subsection{Supersymmetric seesaw type II}
\label{sec:modelII}

In the type II seesaw, neutrino mass generation is triggered by the tree-level exchange of scalar triplets. Its simplest SUSY version requires the addition of a vector-like pair of SU(2) triplet superfields $T$ and $\overline{T}$ of hypercharge $Y=\pm 2$. A natural way to implement the type II seesaw in a GUT scenario is to embed the triplets in a $15$ and $\overline{15}$-plet of SU(5) which decompose under
${\rm SU(3)}\times {\rm SU(2)} \times {\rm U(1)}$ in the following way~\cite{Rossi:2002zb}
\begin{eqnarray}\label{eq:15}
15 & = &  S + T + Z  \thickspace,\\ \nonumber
S & \sim  & (6,1,-2/3), \hskip10mm
T \sim (1,3,1), \hskip10mm
Z \sim (3,2,1/6),
\end{eqnarray}
with an obvious decomposition for the $\overline{15}$. The SU(5) invariant superpotential reads
\begin{eqnarray}\label{eq:pot15}
W & = & \frac{1}{\sqrt{2}} \bY_{15}\, {\bar 5} \, 15 \, {\bar 5}
   + \frac{1}{\sqrt{2}}\lambda_1\, {\bar 5}_H \, 15\,  {\bar 5}_H
+ \frac{1}{\sqrt{2}}\lambda_2\, 5_H \, \overline{15} \, 5_H
+ \bY_5 10\, \, {\bar 5} \, {\bar 5}_H \nonumber \\
 & + & \bY_{10}\, 10 \,10 \, 5_H + M_{15} 15\,  \overline{15}
+ M_5\, {\bar 5}_H \, 5_H \, ,
\end{eqnarray}
with ${5}_H =(H^c,H_u)$ and
${\bar 5}_H=({\bar H}^c,H_d)$. We do not go through the details of the SU(5) breaking as we take the above SU(5) realization only
as a guideline to fix some of the boundary conditions at $M_{GUT}$.
Below $M_{GUT}$, in the SU(5)-broken phase, the superpotential reads
\begin{eqnarray}\label{eq:broken}
W_{II} & = & W_{MSSM} + \frac{1}{\sqrt{2}}(\bY_T L T  L
+  \bY_S D^c S D^c)
+ \bY_Z D^c Z L   \nonumber \\
& + & \frac{1}{\sqrt{2}}(\lambda_1 H_d T   H_d
+\lambda_2  H_u \overline{T}  H_u)
+ M_T T \overline{T}
+ M_Z Z \overline{Z} + M_S S \overline{S}\,.
\end{eqnarray}
The dimension five effective neutrino mass originates now from the decoupling of the triplet states, leading to an effective neutrino mass matrix
\begin{eqnarray}\label{eq:ssII}
\bmm_\nu=\frac{v_u^2}{2} \frac{\lambda_2}{M_T}\bY_T\,,
\end{eqnarray}
once electroweak symmetry is spontaneously broken. It is apparent from the above equation that the flavour structure of $\bmm_\nu$ at low energies is the same as the one of the couplings $\bY_T$ at the decoupling scale $M_T$ (up to renormalization group effects which can be relevant in some special cases~\cite{Joaquim:2009vp}). Consequently, $\bY_T$ is diagonalized by the same matrix as $\bmm_{\nu}$, i.e.
\begin{equation}\label{diagYT}
{\bf {\hat Y}}_T = \bU^T \bY_T\, \bU \,.
\end{equation}
In short, if all neutrino eigenvalues, angles and phases were known, $\bY_T$
would be fixed up to an overall constant which can be easily
estimated to be
\begin{equation}\label{est}
\frac{M_T}{\lambda_2} \simeq 10^{15} {\rm GeV} \hskip2mm
\left(\frac{0.05 \hskip1mm {\rm eV}}{m_{\nu}}\right).
\end{equation}
In principle, the remaining {\em flavoured} Yukawa couplings $\bY_S$ and $\bY_Z$ are not determined by any low-energy neutrino data. Still, they both induce LFV slepton mass terms, just as $\bY_T$ does. Having the above SU(5) GUT model in mind, we impose the unification condition $\bY_T=\bY_S=\bY_Z$ at $M_{GUT}$ in our numerical analysis presented in Section~\ref{sec:num}. As for the heavy-state masses, the mass equality condition $M_T = M_Z = M_Z = M_{15}$ imposed at the GUT scale is spoiled by the renormalization group (RG) running of the masses. Nevertheless, these effects are small and, therefore, gauge coupling unification is maintained. In view of this, for practical purposes we decouple the $T$, $Z$ and $S$ states at the common scale $M_T(M_T)$, neglecting in this way threshold effects resulting from the small RG-induced splittings among the heavy masses.

\subsection{Supersymmetric seesaw type III}
\label{sec:modelIII}

In the case of a type III seesaw model, neutrino masses are generated by the tree-level exchange of zero hypercharge fermions, usually denoted as $\Sigma$,
belonging to the adjoint representation of SU(2). These states can be accommodated, for instance, in a 24-plet of SU(5)~\cite{Perez:2007iw}.
Above the SU(5) breaking scale, the relevant superpotential for our discussion is
\begin{eqnarray}\label{eq:spot5}
W & = & \sqrt{2} \, \bY_5 {\bar 5}_M 10_M {\bar 5}_H
          - \frac{1}{4} \bY_{10} 10_M 10_M 5_H
  +  \bY_{24} 5_H 24_M {\bar 5}_M +\frac{1}{2} \bM_{24} 24_M 24_M \,.
\end{eqnarray}
As in the type II case, we do not specify the Higgs sector responsible for the
SU(5) breaking. The superpotential terms directly involved in neutrino mass generation are those containing the representations $24_M$, which decompose under  ${\rm SU(3)}\times {\rm SU(2)} \times {\rm U(1)}$ as
\begin{eqnarray}\label{eq:def24}
24_M & = &(1,1,0) + (8,1,0) + (1,3,0) + (3,2,-5/6) + (3^*,2,5/6) \thickspace, \\ \nn
   & = & N^c + G + \Sigma + X + \bar{X} \,.
\end{eqnarray}
The fermionic components of $(1,1,0)$ and $(1,3,0)$ have the same quantum numbers as $N^c$ (the type I heavy neutrino singlets) and $\Sigma$. Thus, one expects that, in general, the decoupling of the $24_M$ components leads to an effective neutrino mass operator which contains both a type I and a type III seesaw contribution. In the SU(5) broken phase the superpotential is
\begin{eqnarray}\label{eq:spotIII}
 W_{III} & = &  W_{MSSM}
 +  H_u\left( \bY_\Sigma \Sigma - \sqrt{\frac{3}{10}}
               \bY_\nu N^c \right) L
 + \bY_X H_u \bar X D^c \nonumber \\
         & & + \frac{1}{2} \bM_{R} N^c N^c
         + \frac{1}{2} \bM_{G} G G
          + \frac{1}{2} \bM_{\Sigma} \Sigma \Sigma
          + \bM_{X} X \bar X\,.
\end{eqnarray}
Once more, we impose the GUT scale boundary condition
$\bY_\Sigma = \bY_\nu = \bY_X$ and $\bM_R = \bM_G=\bM_\Sigma=\bM_X$. Integrating out the heavy fields, and after EWSB,
the following effective neutrino mass matrix is generated:
\begin{equation}
\bmm_\nu = - \frac{v^2_u}{2}
\left( \frac{3}{10} \bY^T_\nu \bM^{-1}_R \bY_\nu + \frac{1}{2} \bY^T_\Sigma \bM^{-1}_\Sigma \bY_\Sigma \right).
\label{eq:mnu_seesawIII}
\end{equation}
As mentioned above, there are two contributions to neutrino masses stemming from the gauge singlets $N^c$ as well as from the SU(2) triplets $\Sigma$. In this case the extraction of the Yukawa couplings from low-energy parameters for a given high scale spectrum is more complicated than in the other two types of seesaw models. However, as we start from universal couplings and masses at $M_{GUT}$, we find that at the seesaw scale one still has $\bM_R \simeq \bM_\Sigma$ and $\bY_\nu \simeq \bY_\Sigma$. Consequently, one has
\begin{equation}
\bmm_\nu \simeq - v^2_u  \frac{4}{10} \bY^T_\Sigma \bM^{-1}_\Sigma \bY_\Sigma\,,
\label{eq:mnu_seesawIIIa}
\end{equation}
to a good approximation. This result allows us to use the same decomposition for $\bY_\Sigma$ as the one discussed in section \ref{sec:modelI}, up to the overall factor $4/5$ [see Eq.~(\ref{Ynu})].

\section{Lepton flavour violation in the (s)lepton sector}
\label{sec:LFV}

The search for LFV processes beyond neutrino oscillations has attracted a great deal of attention both from the experimental and theoretical communities. Rare decays like $\mu \to e \gamma$ have been searched for decades, without any positive result. The most stringent constraint on this process comes from the MEG experiment~\cite{meg} which, by analysing the data collected in 2009 and 2010~\cite{Adam:2011ch}, has set the new bound $\text{Br}(\mu \to e
\gamma) < 2.4 \cdot 10^{-12}$.

The branching ratio (BR) for $l_i \to l_j \gamma$ can be generically written as
\cite{Kuno:1999jp}
\begin{equation} \label{brLLG}
\text{Br}(l_i \to l_j \gamma) = \frac{48 \pi^3 \alpha}{G_F^2} \left(
|\bA_L^{ij}|^2 + |\bA_R^{ij}|^2 \right) \text{Br}(l_i \to l_j \nu_i \bar{\nu}_j)
\,.
\end{equation}
The amplitudes $\bA_L$ and $\bA_R$ depend on the specific physics framework and, in general, are generated at the 1-loop level.
In our SUSY scenario, the dependence of those amplitudes on the LFV slepton soft masses is approximately given by
\begin{equation} \label{A-dependence}
\bA_L^{ij} \sim \frac{(\bmm_{\tilde L}^2)_{ij}}{m_{SUSY}^4} \quad , \quad \bA_R^{ij} \sim \frac{(\bmm_{\tilde{e}^c}^2)_{ij}}{m_{SUSY}^4} \,,
\end{equation}
where $\bmm_{\tilde L}^2$ and $\bmm_{\tilde{e}^c}^2$ are the doublet and singlet slepton soft mass matrices, respectively, and $m_{SUSY}$ is a typical supersymmetric mass. In the derivation of these estimates one typically assumes that (a) chargino/neutralino
masses are similar to slepton masses and (b) left-right flavour mixing induced by $A$-terms is negligible\footnote{This assumption is not valid when large values of $|A_0|$ are considered. Nevertheless, the above estimates can be still used to illustrate the dependence of the BRs on the low-energy neutrino parameters.}.

Assuming universal boundary conditions for the soft SUSY-breaking terms at the GUT scale, and considering only the leading-log approximation for the LFV slepton masses and trilinear terms induced through RG running, one obtains:
\begin{eqnarray}
(\bmm_{\tilde L}^2)_{ij} &\simeq& -\frac{a_k}{8 \pi^2 }
 \left( 3 m^2_0 +  A^2_0 \right)
 \left(\bY^{\dagger}_k \bL \bY_k\right)_{ij} \thickspace, \\
(\bA_{e})_{ij} &\simeq& -a_k \frac{3}{ 16 \pi^2 }   A_0
 \left(\bY_e \bY^{\dagger}_k \bL \bY_k\right)_{ij} \thickspace,
\label{eq:LFVentriesI}
\end{eqnarray}
for $i\ne j$. In the basis where $\bY_e$ is diagonal,
$\bL_{mn} = \ln(M_{GUT}/M_n)\delta_{mn}$ and $\bY_k$ is the Yukawa
coupling of the type-$k$ seesaw ($k=I,II,III$) with $\bY_k=(\bY_\nu,\bY_T,\bY_\Sigma)$, given at $M_{GUT}$. Taking into account the renormalisation group equations (RGEs) for $\bmm_{\tilde L}^2$ and $\bA_e$ we obtain
\begin{equation}
a_I=1\,\, , \,\, a_{II}=6 \,\, \mathrm{and} \,\,\, a_{III} = 9/5 \, .
\label{eq:LFVentriesII}
\end{equation}
Note, that in case of the type II seesaw the matrix $\bL$ is
proportional to the identity and thus can be factored out. All models considered here
have in common that they predict negligible flavour violation for the
RH sleptons
\begin{eqnarray}
(\bmm_{\tilde{e}^c}^2)_{ij} &\simeq& 0.
\end{eqnarray}
Although not very accurate, the above approximations allow to estimate the LFV slepton masses and $A$-terms within different seesaw frameworks. The BRs for rare lepton decays $l_i \to l_j \gamma$ are roughly given by
\begin{equation}
{\rm Br}(l_i \to l_j \gamma) \propto \alpha^3 m_{l_i}^5
	\frac{|(\bmm_{\tilde L}^2)_{ij}|^2}{m_{SUSY}^8}\tan^2\beta.
\label{eq:LLGapprox}	
\end{equation}
For distinct seesaw scenarios, and a given set of high-scale parameters, the above BRs change due to the different $(\bmm_{\tilde L})_{ij}^2$ and the distorted mass spectrum (which differs from the pure CMSSM one). The most important parameter turns out to be the seesaw scale due to its influence on the size of the Yukawas. The higher the seesaw scale is, the larger are the Yukawa couplings and, consequently, the LFV rates. In case of the type II seesaw, the coupling $\lambda_2$ plays a crucial r\^{o}le, as seen in Eq.~\eqref{est}. Small values of this parameter lead to large $\bY_T$ Yukawa couplings and high LFV rates.

Finally, we would like to comment on the influence of the $\bR$ matrix
on LFV decay rates. As shown in Eq.~\eqref{Ynu}, the $\bY_\nu$ Yukawa
couplings for type I seesaw are proportional to $\bR$ and, thus,
different choices of this matrix lead to different off-diagonal
entries in the soft squared mass terms (which in turn changes the LFV
rates). Similarly, the type III Yukawa couplings, $\bY_\Sigma$, follow
an analogous equation and, consequently, also change with $\bR$. This
additional freedom can be used to cancel some $\left(\bY^{\dagger}_k
\bL \bY_k\right)_{ij}$ combinations, in particular the one with
$(i,j)=(\mu,e)$~\cite{Casas:2001sr}. This allows for large LFV effects
in the $\tau-e$ and $\tau-\mu$ sectors while having negligible $\mu-e$
transitions. In the following, we will disregard this
possibility\footnote{In fact, we will always consider real parameters
and degenerate spectra for the right-handed (RH) neutrinos (in type I)
and for the SU(2) fermion triplets (in type III). In such scenarios
the $\bR$ matrix is physically irrelevant, since it drops out in the
computation of $\left(\bY^{\dagger}_k \bL
\bY_k\right)_{ij}$~\cite{Casas:2001sr}. For a discussion on the
effects of considering complex parameters we address the reader to,
e.g. Refs.~\cite{Raidal:2008jk,Branco:2011zb}.}. Therefore,
implications on $M_{SS}$ drawn from $\mu \to e \gamma$ considerations
can be regarded as approximate lower bounds\footnote{Once
Br$(\mu \to e \gamma)$ and $m_{\tilde{\nu}} \sim m_{SUSY}$ are known,
one can determine $M_{SS}$ assuming $\bR={\bf 1}$. Under this assumption, an upper limit
on Br$(\mu \to e \gamma)$ can lead to an upper limit on $M_{SS}$, once $m_{\tilde{\nu}} \sim m_{SUSY}$ is (at least
approximately) known. Larger $M_{SS}$ are in principle possible if
$\bR$ is tuned to obtain a cancellation in the
$\mu-e$ sector. However, one cannot find $\bR$ matrices that allow to
go to much smaller $M_{SS}$ scales.}.

\section{Numerical analysis and results}
\label{sec:num}

\subsection{Setup}
\label{subsec:numsetup}

Our numerical results have been obtained with {\tt SPheno}
\cite{Porod:2003um,Porod:2011nf}. Taking as input the SM parameters, as
well as the usual universal soft terms at the GUT scale
\begin{equation}
m_0 , M_{1/2} , A_0 , \tan \beta , \text{sign}(\mu) ,
\end{equation}
{\tt SPheno} computes the resulting SUSY spectrum by means of complete
2-loop RGEs \cite{Martin:1993zk,Yamada:1994id,Jack:1994kd}, properly
adapted for every model. This includes the pure CMSSM and the three
seesaw variants studied in this work. At the SUSY scale, the $\mu$
parameter is obtained including the most relevant 2-loop
corrections~\cite{Dedes:2002dy} and complete 1-loop corrections to all
sparticle masses are implemented \cite{Pierce:1996zz}. These
calculations follow the $\overline{DR}$ renormalization scheme.

In case of the Higgs boson mass, the aforementioned 1-loop corrections
are supplemented by the most relevant
$\mathcal{O}[\alpha_s(\alpha_t+\alpha_b)+(\alpha_t+\alpha_b)^2+\alpha_\tau
\alpha_b + \alpha_\tau^2]$ 2-loop contributions
\cite{Degrassi:2001yf,Brignole:2001jy,Brignole:2002bz,Dedes:2003km,
Dedes:2002dy,Allanach:2004rh}. For a detailed study of the {\tt
SPheno} results for the Higgs boson mass and a comparison to other
popular numerical codes we refer to \cite{Allanach:2004rh}. We have
checked that our results agree, within the usual $2-3~\GeV$
theoretical uncertainty, with the results given by {\tt
FeynHiggs}~\cite{Heinemeyer:1998yj}. This code uses an on-shell
renormalization scheme and therefore small differences are expected on
theoretical grounds. In particular, larger differences are found for
very large Higgs boson masses, $m_{h^0} \sim 129-130~\GeV$, a region
where numerical computations are no longer accurate. 

Uncertainties in the Higgs mass calculation have been often discussed
in the literature. In short, the dominant sources
of the theoretical error on $m_{h^0}$ are the uncertainty in the top (bottom) mass,
the missing (sub-dominant) 2-loop contributions and the missing dominant 3-loop
diagrams in public codes. Currently, the Particle Data Group quotes $m_{t} =
173.5 \pm 1.0$~\cite{Beringer:1900zz}, leading to
$\Delta m_{h^0} \lsim 1$ GeV, depending on the parameter point. We note
in passing that a complete 2-loop calculation based on the
Higgs effective potential exists in the literature \cite{Martin:2002wn}. Moreover, 
3-loop contributions to the Higgs mass have also been calculated~\cite{Martin:2007pg,Kant:2010tf}. 
So far, none of these contributions~\cite{Martin:2002wn,Martin:2007pg,Kant:2010tf} 
have been implemented into a public code.

We also found good agreement between our results and those presented
in some recent works devoted to the study of the Higgs mass in the
MSSM~\cite{Arbey:2011ab,Baer:2011ab,Ellis:2012aa,Brummer:2012ns}. Although
the theoretical error is always present, and the exact numbers might
differ in some cases, the general behaviour and the dependence on the
SUSY parameters are correctly reproduced. We have decided not to
compute the SUSY spectrum at a fixed scale $Q = 1~\TeV$, as suggested
by the SPA conventions \cite{AguilarSaavedra:2005pw}, since that is
known to give a poor accuracy in the determination of the Higgs boson
mass for scenarios with very large values of $A_0$ or with multi-TeV
stop masses.  Instead, we compute the SUSY spectrum at the geometric
average of the two stop masses $m_{\tilde t_{1,2}}$, i.e. $M_S =
\sqrt{ m_{\tilde t_1} m_{\tilde t_2}}$.

Although we evaluate the Higgs mass numerically taking into account the
higher-order corrections enumerated above, we find it useful to recall
that the leading 1-loop corrections to the Higgs mass for moderate
values of $\tan\beta$ and large Higgs pseudoscalar mass $m_A$, are
approximately given
by~\cite{Okada:1990vk,Ellis:1990nz,Haber:1990aw,Carena:1995bx}
\begin{equation}
\label{eq:mhapp}
m_{h^0}\simeq m_Z^2\cos^2\beta + \frac{3m_t^4}{4\pi^2v^2}\left[\ln\left(\frac{M_S^2}{m_t^2}\right)+
\frac{X_t^2}{M_S^2}\left(1-\frac{X_t^2}{12M_S^2 }\right)\right]\;,\;X_t = A_t -\mu \cot\beta\,,
\end{equation}
where $\mu$ is the Higssino mass parameters, $A_t$ is the top
trilinear term at low-energy and $X_t$ is the mixing parameter in the
stop sector. Obviously, the above approximation is not always very
accurate. In any case, we will only use it to understand the behaviour
of the Higgs mass with some of the input parameters of the seesaw
models discussed in the previous section.

In the following we will present and discuss our numerical
results. Notice that we will loosely talk about ``the seesaw scale'',
$M_{SS}$, when referring to the mass of the seesaw mediators, i.e. the
right-handed neutrino mass, $M_R$, in case of seesaw type I, the $Y=2$ triplet
mass, $M_{15}$ (or $M_T$), for type II or the mass of the $Y=0$
triplet, $M_{24}$ (or $M_{\Sigma}$), for type III. Our assumptions
regarding the input parameters for each of the seesaw models
are:\medskip

{\bf Type I:} We consider the general case of 3 degenerate RH
neutrinos with mass $M_R$. In the flavour sector, we fix $\bR={\bf 1}$
[see Eq.~(\ref{Ynu})]. This choice does not have a significant impact
on the Higgs mass since, as already pointed out, the effect of the
Yukawa couplings on $m_{h^0}$ is marginal. As shown below, 
even the model with three copies of degenerate RH neutrinos is always 
very close to the CMSSM limit. Therefore, we will not discuss variants 
with less RH neutrinos or with non-degenerate masses.
\medskip

{\bf Type II:} Apart from the unification conditions for the Yukawa
couplings and masses of the different 15-plet components mentioned in
Section~\ref{sec:modelII}, we will use in most of the cases the values
$\lambda_{1,2}(M_{GUT})=0.5$ for the superpotential couplings of the
triplets with the Higgs superfields. Later, we will comment on how
relaxing this condition affects the Higgs and squark masses. \medskip

{\bf Type III:} We will always assume the existence of three copies of
24-plets, with an approximately degenerate mass
$M_{\Sigma}$. Alternatively, one could also explain neutrino data with
two degenerate 24's or with three, being one ``light'' and the other
two close to the GUT scale. The first of these options leads to
results somewhere between those shown for type II and type III with
three degenerate 24's, while the latter has $m_{h^0}$ somewhere
between type II and type I. Since nothing qualitatively new results
from these cases, we will not discuss them in detail. As in the type I
case we assume $\bR={\rm {\bf 1}}$.\medskip

For all our numerical cases the values of the low-energy neutrino
parameters (mixing angles and mass-squared differences) coincide with
the best-fit values provided by global analysis of all neutrino
oscillation data~\cite{Tortola:2012te,Fogli:2012ua,SchwetzTalk}.  To
simplify our analysis, we consider all couplings and mass parameters
to be real and for $\tan\beta$ we take the reference value
$\tan\beta=20$.  For other values of $\tan\beta$, our CMSSM results
agree quite well with those discussed, for example, in
\cite{Arbey:2011ab}. We have scanned the parameters $m_0$ and
$M_{1/2}$ in the range of $[0,10]\,\TeV$.  As for $A_0$, we have taken
values in the interval $[-5,5]\,\TeV$, although we will mainly
concentrate on the two extreme cases with $A_0=0\,\TeV$ and
$A_0=-5\,\TeV$. For other choices of $A_0$ (and $\tan\beta$)
the results always lie between the extreme ones, as 
discussed in detail for the CMSSM in Refs.~\cite{Arbey:2011ab,Brummer:2012ns}. 
Since our findings agree with these works, we do not repeat the discussion here.

Current bounds on squark and gluino masses in CMSSM-like setups
from  ATLAS \cite{Aad:2012hm} and CMS \cite{Cha:2012mf} already
exclude $m_{\tilde g} =m_{\tilde q} \simeq 1.4$ TeV and
$m_{\tilde g} \simeq (800-900)$ GeV for very heavy squarks.
Therefore, we will mainly concentrate on parts of the parameter
space where $m_{\tilde g}$ and $m_{\tilde q}$ are larger than
1 TeV.

There are several other constraints on SUSY from different
searches in the literature. However, as shown below, our spectra
are always relatively heavy and, therefore, they pass all other known experimental
constraints (once we impose the Higgs mass window).
Of particular importance is the recent upper limit on $B^0_s \to \mu^+\mu^-$~\cite{Eerola:2012yg}, which 
particularly constrains the large $\tan\beta$ region of the
SUSY parameter space~\cite{Buchmueller:2012hv}. Since in our numerical examples we use the moderate value $\tan\beta=20$, the $B^0_s \to \mu^+\mu^-$ bound is not exceeded.

\subsection{Results}
\label{sec:results}

It is well known that adding seesaw mediators with masses between the
SUSY and GUT scales changes the RG running of gauge couplings. As a
result, the RG flow of all Yukawa couplings and mass parameters is
modified with respect to the CMSSM
case~\cite{Hirsch:2008gh,Esteves:2009qr,Esteves:2010si,
Esteves:2011gk,Biggio:2012wx}. In the case of type II and III seesaws,
the increase in the value of the common gauge coupling
$\alpha(M_{GUT})$ leads, in general, to lighter sparticles
\cite{Hirsch:2008gh,Esteves:2009qr}. Therefore, one expects the Higgs
mass to be sensitive to the parameters characterising each seesaw
model, namely the mass $M_{SS}$ and possible couplings with the Higgs
and/or lepton sectors of the MSSM. Consequently, the reconstruction of
the SUSY-breaking parameters at the universality scale $M_{GUT}$ from
low-energy mass measurements will be very sensitive to the presence of
new fields at intermediate scales.

\begin{figure}
\includegraphics[width=74mm]{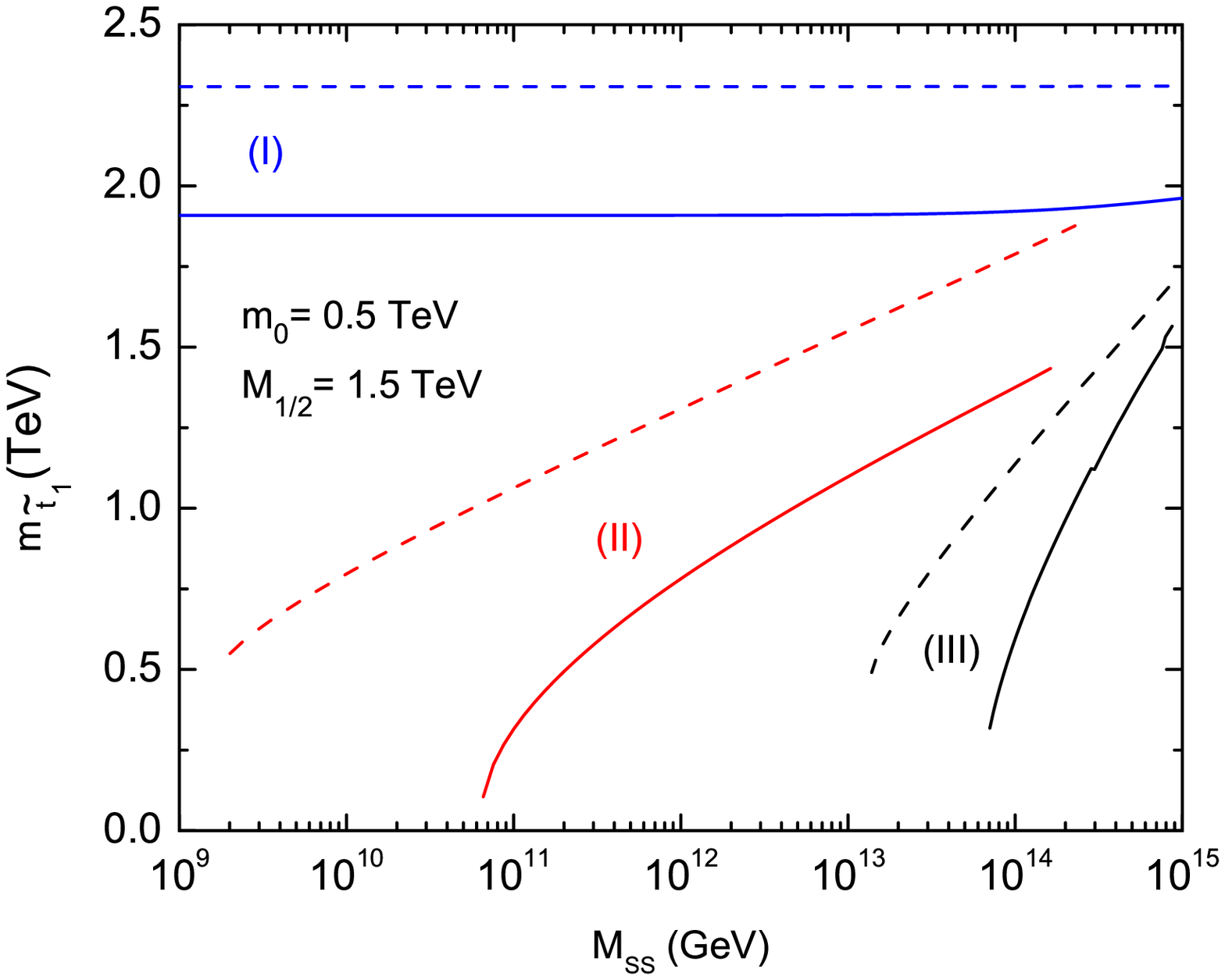}
\includegraphics[width=74mm]{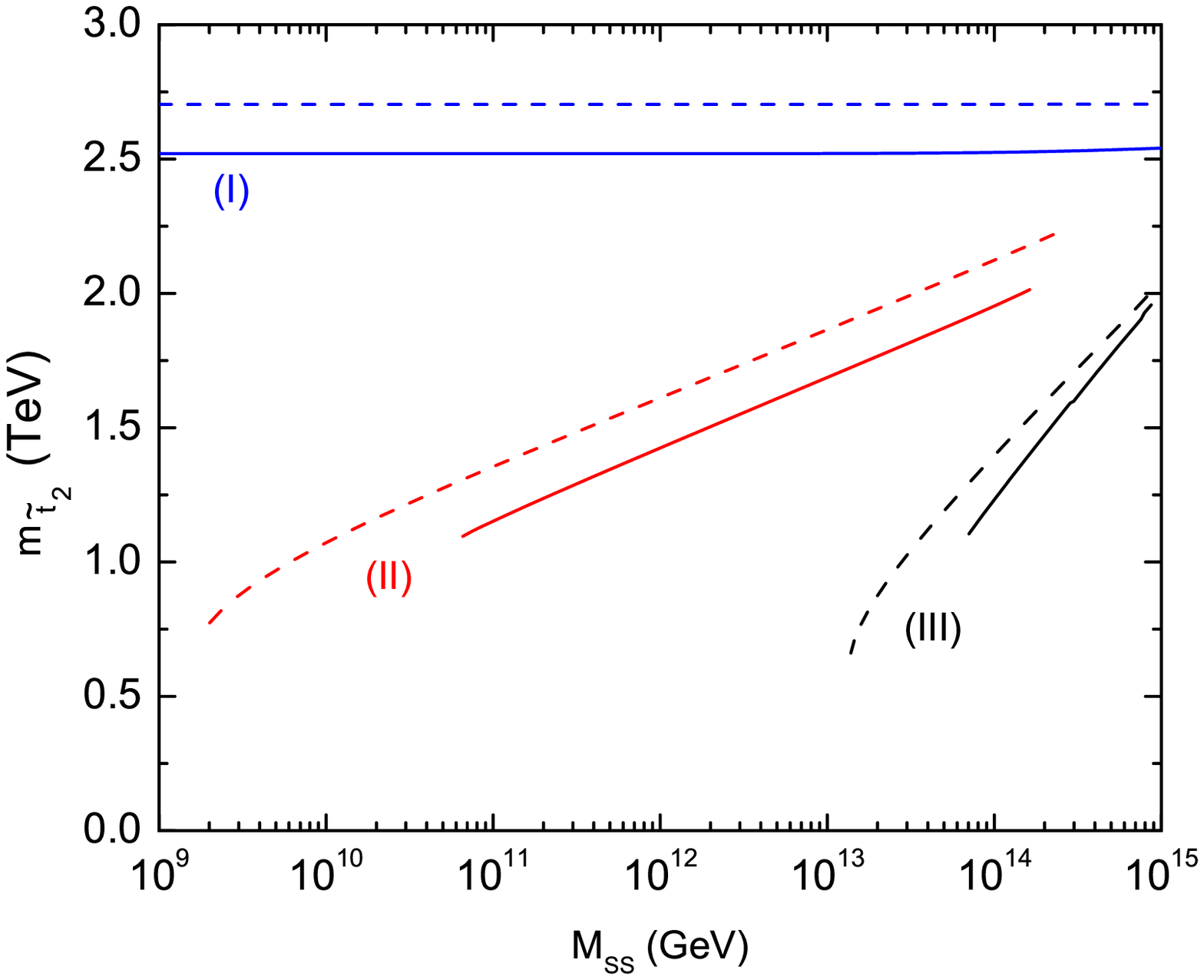}
\includegraphics[width=74mm]{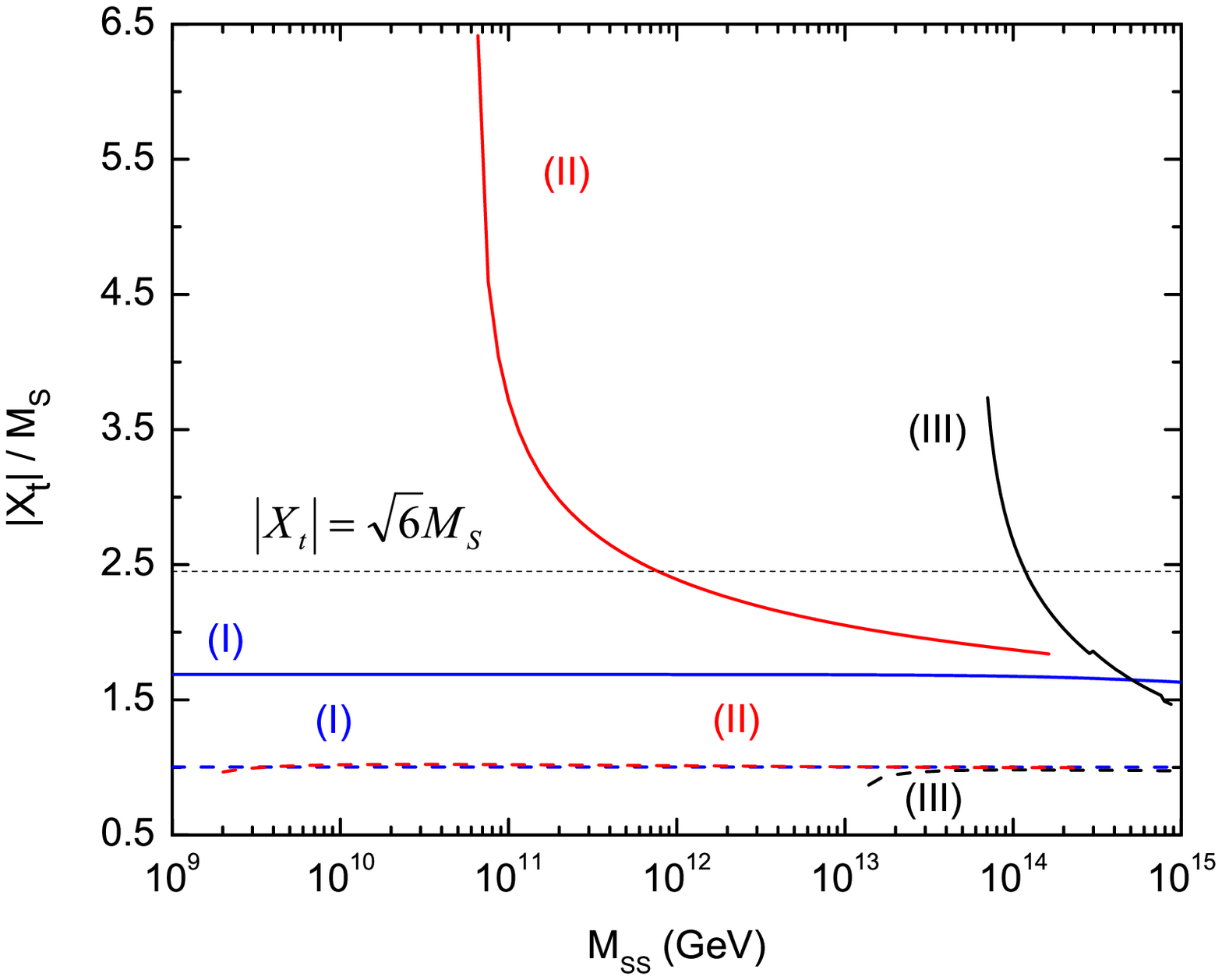}
\hspace*{2mm}\includegraphics[width=74mm]{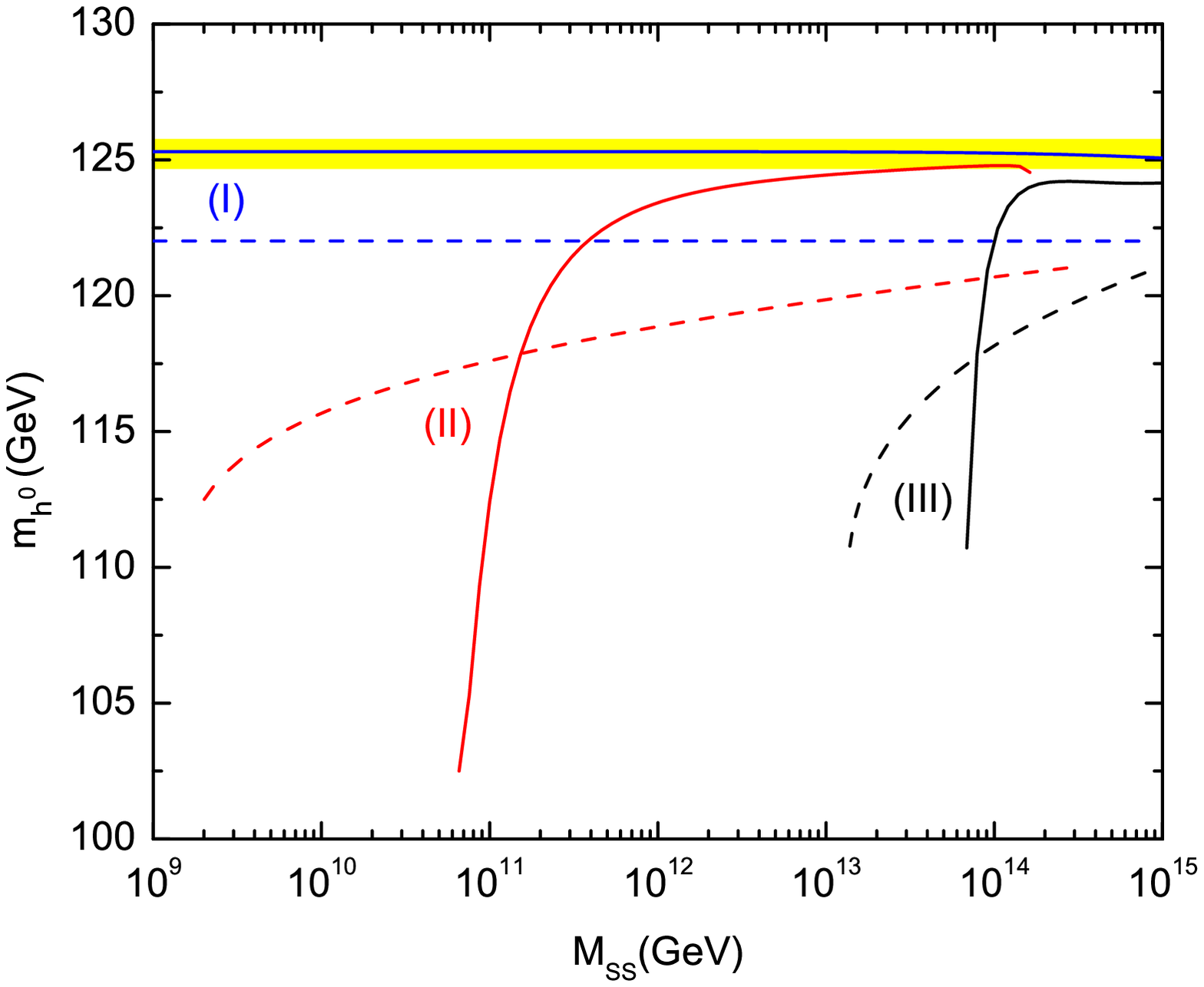}
\caption{\label{fig:exa} Variation of the scalar top masses (top
panels), the ratio $|X_t/M_S|$ and the mass of the lightest Higgs
$h^0$ (bottom panels) with $M_{SS}$, for a particular point in the
SUSY parameter space with $m_0=0.5~\TeV$, $M_{1/2}=1.5~\TeV$. The blue,
red and black lines correspond to seesaw type I, II and III,
respectively. The results are shown for $A_0=0~\TeV$ (dashed lines) and
$A_0=-3~\TeV$ (solid lines). For values of the seesaw scale larger than
roughly $10^{15}~\GeV$ no solutions consistent with observed neutrino
data can be found. Also, for $M_{SS} \lesssim$ (few) $10^{9}$
[$10^{13}$]~GeV gauge couplings become non-perturbative below the GUT
scale in case of seesaw type II [type III]. See also text.}
\end{figure}

As an example, in Fig.~\ref{fig:exa} we show the behaviour of the stop
masses $m_{\tilde{t}_{1,2}}$ and mixing parameter $X_t$, as well as
the mass of the lightest Higgs, $h^0$, as a function of the seesaw
scale $M_{SS}$, for seesaw type I (blue), type II (red) and type III
(black) taking $m_0=0.5~\TeV$ and $M_{1/2}=1.5~\TeV$. The results are
shown for two values of the common trilinear term at the GUT scale,
namely $A_0=-3~\TeV$ (solid lines) and $A_0=0~\TeV$ (dashed lines). For
values of the seesaw scale larger than roughly $10^{15}~\GeV$ no
solutions consistent with observed neutrino data can be found, while
for values of the seesaw scale below approximately (few) $10^{9}$
$(10^{13})\,\GeV$ gauge couplings become non-perturbative below the GUT
scale in case of seesaw type II (type III). This explains why no
results are shown for lower values of $M_{SS}$ in those cases.

The first immediate (and expected) conclusion that one can infer from
the results presented in this figure is that there is essentially no
dependence of $m_{h^0}$ on $M_{R}$ in case of type I (bottom-right
panel in Fig.~\ref{fig:exa}). This is due to the fact that sparticle
masses do not change with $M_R$, as can be seen for the particular
cases of $m_{\tilde{t}_{1,2}}$ (top panels) and neither stop mixing
does. Due to the singlet nature of the RH neutrinos in the type I
seesaw, and to the fact that they only have Yukawa couplings $\bY_\nu$
with the lepton and Higgs doublet superfields [see
Eq.~(\ref{eq:WtypeI})] the soft SUSY breaking MSSM parameters affected
at the 1-loop level are $m_{H_u}^2$, $\bmm_{\tilde L}^2$, $\bA_{e}$ and
$\bA_{u}$. Still, even those show only very mild departures from their
CMSSM values. All other soft masses change only at the 2-loop level.

The neutrino Yukawas $\bY_\nu$ required to fit neutrino data depend on
$M_{R}$ and are ${\cal O}(1)$ for $M_{R} \simeq 10^{15}~\GeV$.  The
results of all plots in Fig.~\ref{fig:exa} show that, even for such
large Yukawas, the changes of the SUSY spectrum are relatively small,
due to the {\em short} RG running from the GUT scale to $M_{R}$. 
\footnote{A shift of $m_{h^0}$ of the order of several GeV was found
in \cite{Heinemeyer:2010eg} in case of type-I seesaw, if the soft SUSY
breaking mass term $m_M$ for the right-sneutrinos is of the order of
$M_R$.  For $m_M \sim m_{SUSY} \sim {\cal O}({\rm few TeV})$ (as it is in our case), the shift in the Higgs mass
is always less than ${\cal O}(0.1)$ GeV, i.e. far below the theoretical
uncertainty of the calculation.}  For smaller values of $M_{R}$ no
traces of the seesaw remain in the SUSY
spectrum~\cite{Arbelaez:2011bb}. The only important consequence of
changing $M_{R}$ is the strong effect of this scale on the LFV entries
of $\bmm_{\tilde L}^2$ which control the rates of LFV processes like
$\mu\to e \gamma$ (see Section~\ref{sec:LFV} and the discussion
below). Moreover, as the results of Fig.~\ref{fig:exa} show, changing
the value of $A_0$ from zero to -3~TeV, shifts down the stop masses
(middle left panel) due to the term proportional to $A_t^2$ in the RGE
of $(\bmm_{\tilde{u}^c}^2)_{33}$. On the other hand, the magnitude of
the stop mixing parameter $X_t$ increases as a consequence of the fact
that, at low energies, $|A_t|$ is larger for $A_0=-3~\TeV$ than for
$A_0=0~\TeV$. Therefore, $|X_t/M_S|$ increases when going from
vanishing $A_0$ to $A_0=-3~\TeV$, resulting in an increase of the
Higgs mass by approximately $3~\GeV$. Of course, this feature is also
present in the CMSSM and is by no means related with the presence of
the heavy neutrino singlets.

The situation changes when one turns to the type II and type III
seesaws. In these cases, for a fixed choice of CMSSM parameters, one
usually finds that stop masses become smaller when lowering the seesaw
scale, $M_{SS}$. At the same time, the stop mixing angle can
increase. However, this increase is practically never sufficient to
compensate for the smaller stop masses. Thus, in general, $m_{h^0}$
decreases with decreasing $M_{SS}$ for both type II and type III. As
Fig.~\ref{fig:exa} shows this decrease depends also on $A_0$, with
changes in $m_{h^0}$ being much smoother (and smaller) for $A_0=0\,\TeV$
than for $A_0=-3~\TeV$. For the lowest values of $M_{SS}$ possible,
$m_{h^0}$ can be even lighter for $A_0=-3~\TeV$ than for $A_0=0\,\TeV$. This
is due to the rather strong dependence of the stop masses on $A_0$.
All these features, discussed here for a special CMSSM point, are
qualitatively valid for rather larger ranges on the CMSSM parameter
space, as we will discuss next.

\begin{figure}[t]
\begin{tabular}{ll}
\includegraphics[width=72mm]{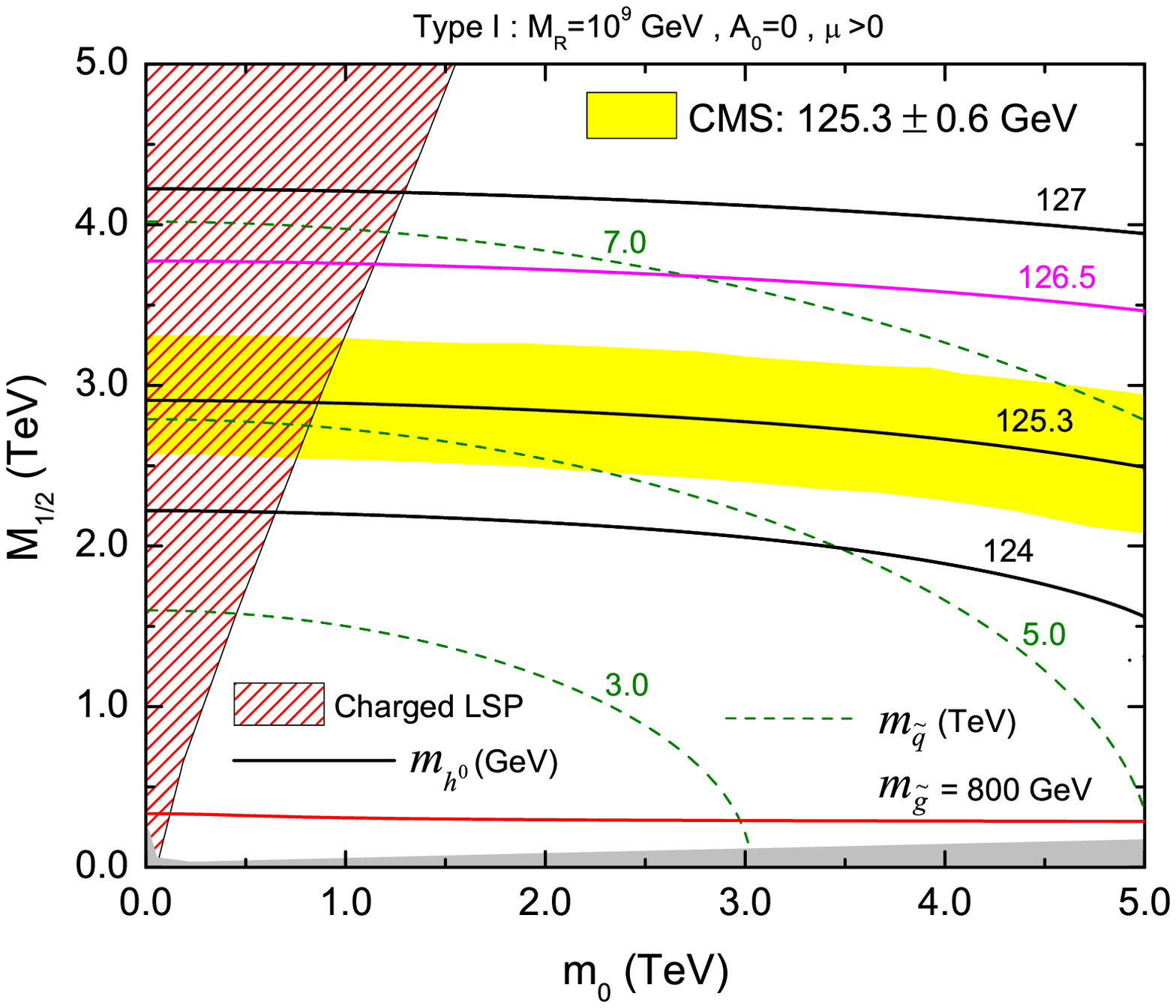} &
\includegraphics[width=72mm]{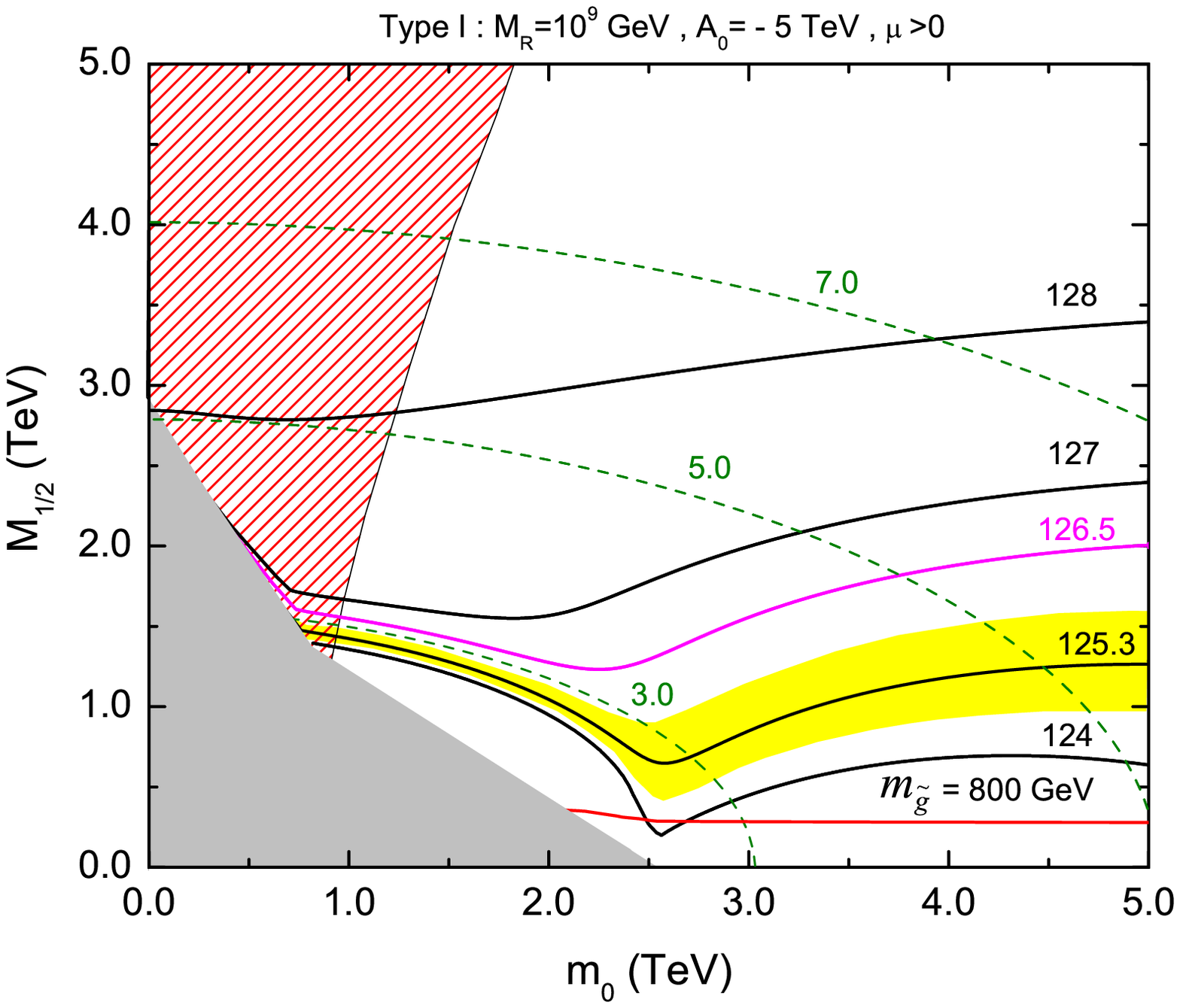}\\
\includegraphics[width=72mm]{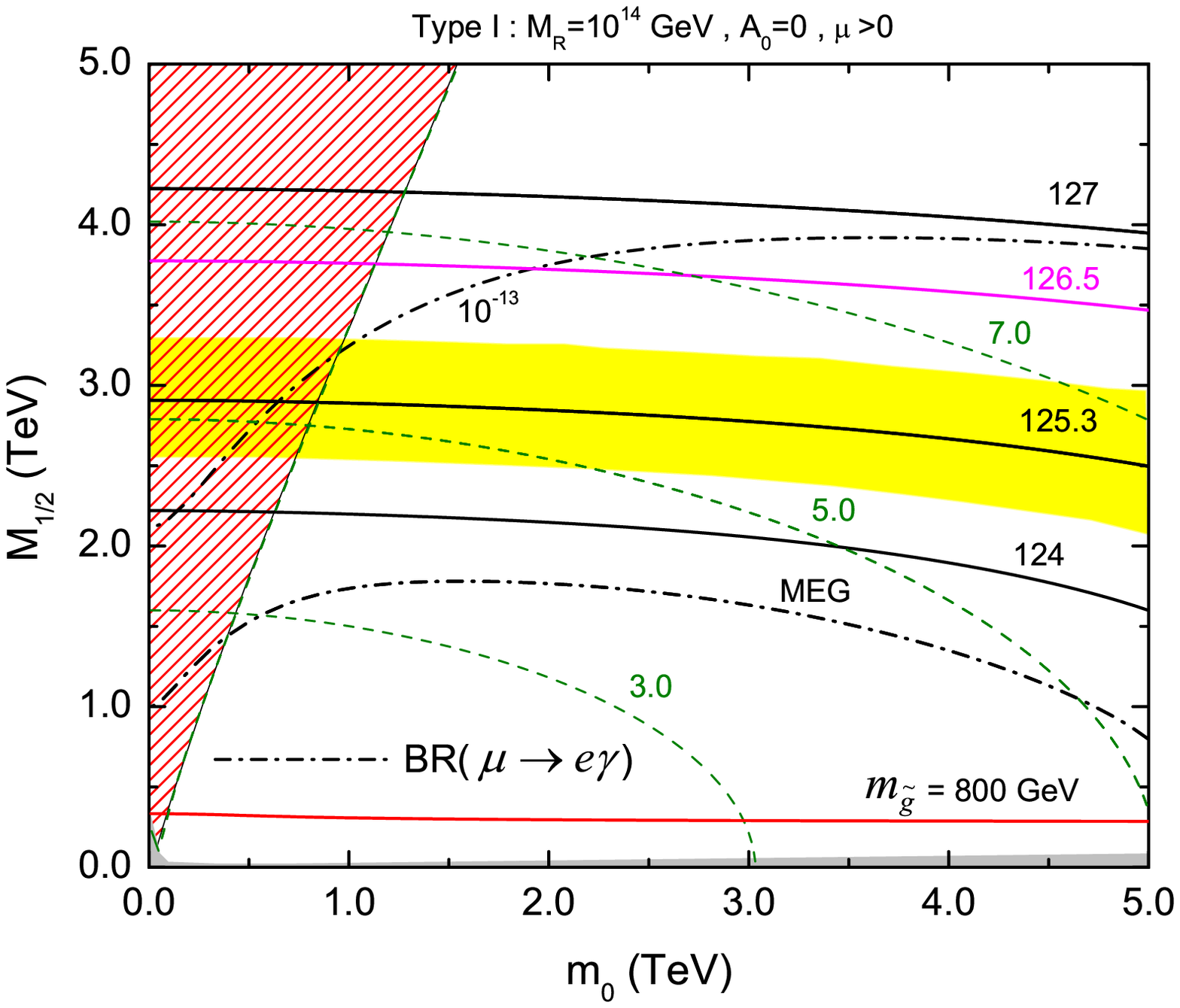}&
\includegraphics[width=72mm]{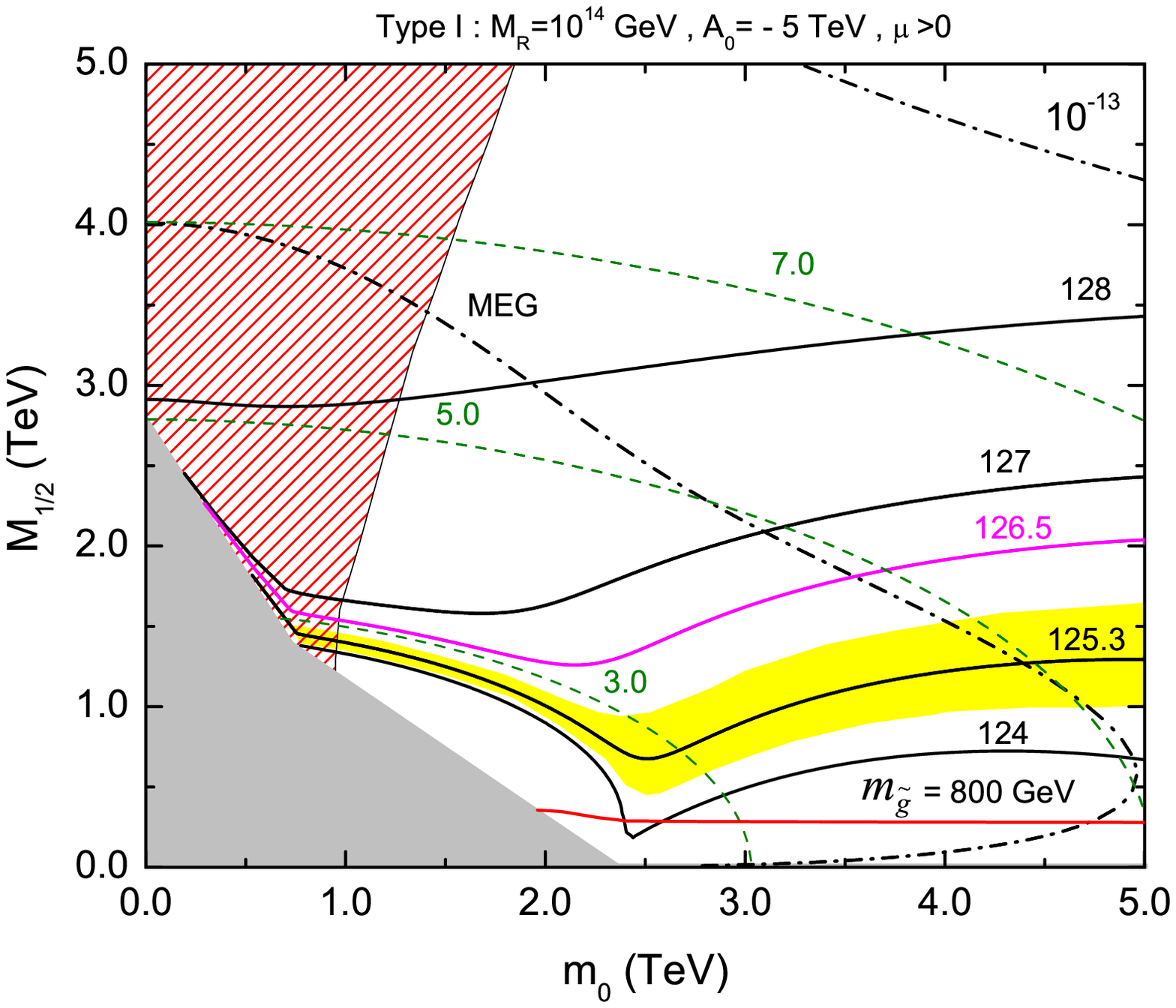}
\end{tabular}
\caption{\label{fig:m0m12typeI}Examples of squark mass, Higgs mass
and Br($\mu\to e \gamma$) contours in the plane ($m_0,M_{1/2}$) for
CMSSM plus seesaw type I for two values of the seesaw scale $M_{R}$:
$M_{R}=10^{9}~\GeV$ (top) and $M_{R}=10^{14}~\GeV$ (bottom); as well
as two values of $A_0$: $A_0=0\,\TeV$ (left) and $A_0=-5~\TeV$ (right).
Br($\mu\to e \gamma$) is orders of magnitude below
the expected experimental sensitivity in case of $M_{R}=10^{9}~\GeV$
and, therefore, contours are not shown (for a discussion see text).}
\end{figure}

\begin{figure}[t]
\begin{tabular}{ll}
\includegraphics[width=72mm]{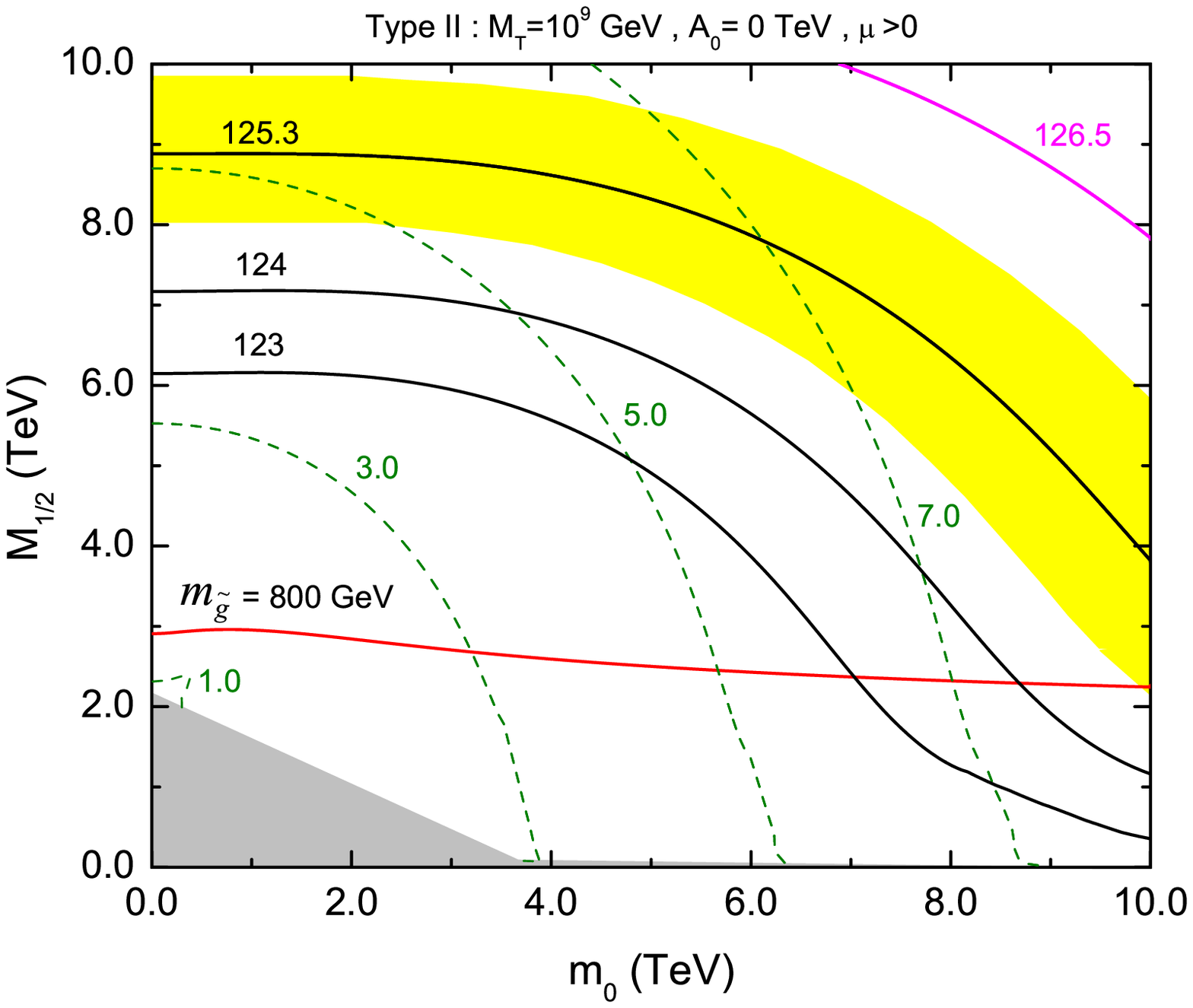} &
\includegraphics[width=72mm]{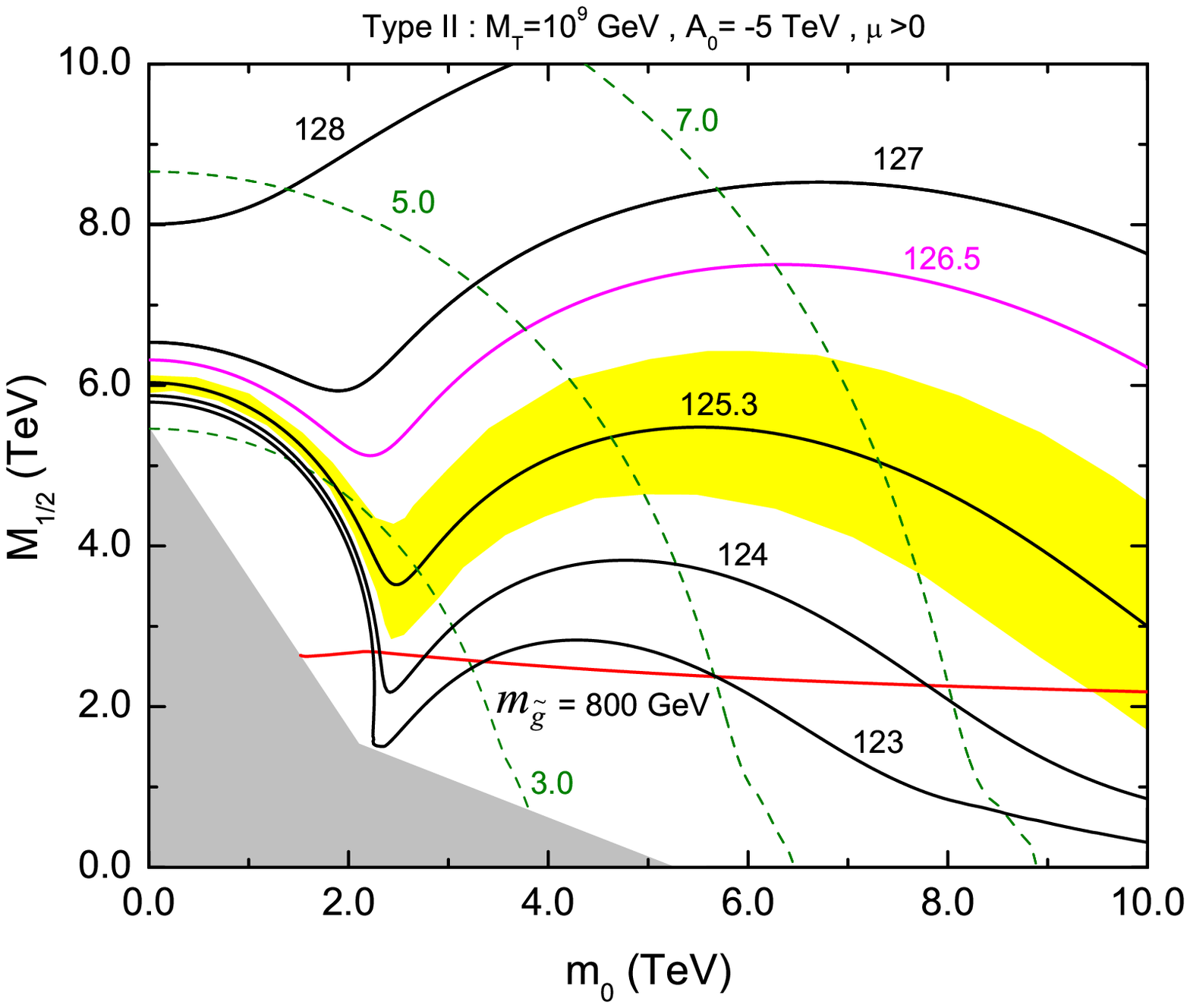}\\
\includegraphics[width=72mm]{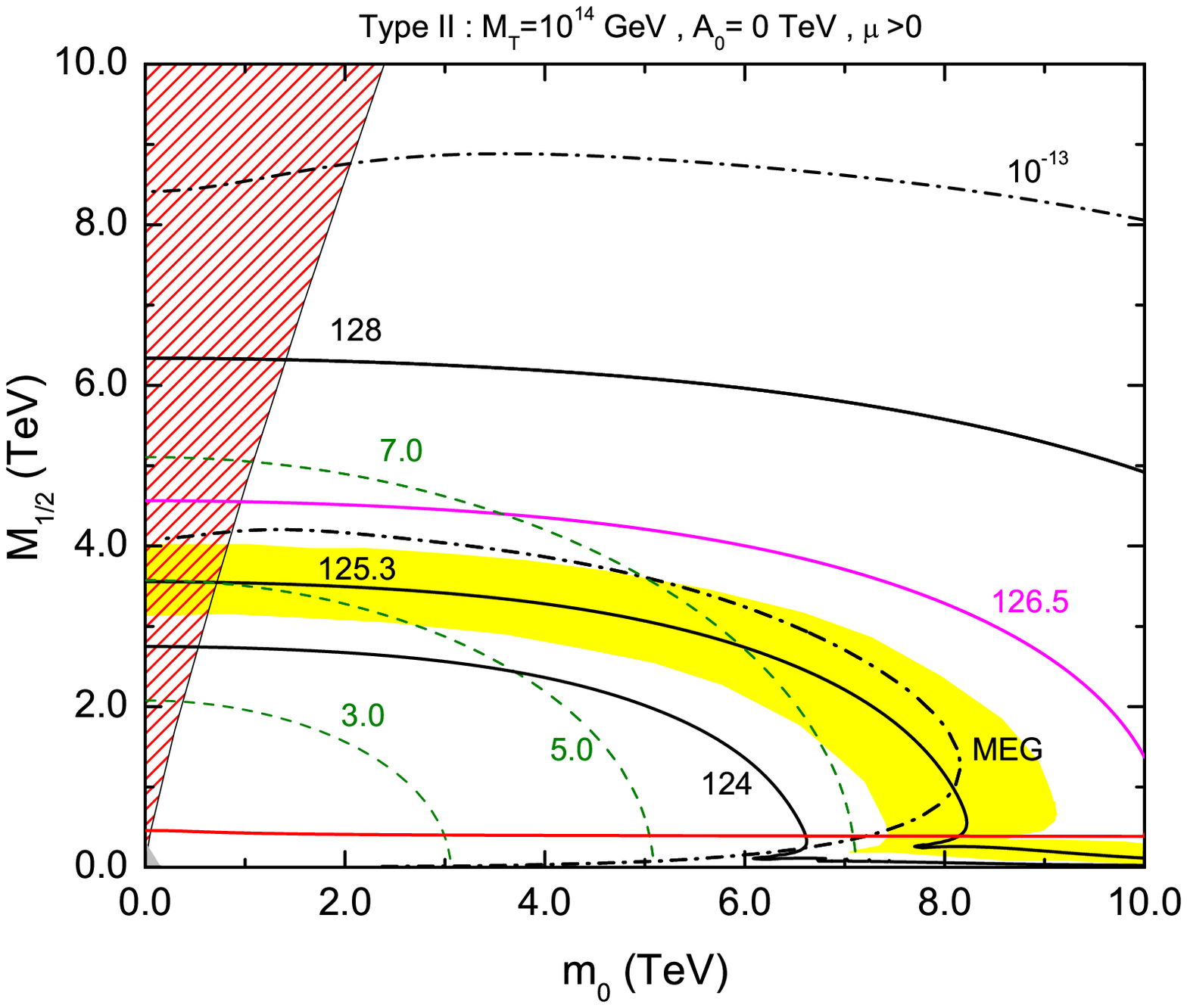} &
\includegraphics[width=72mm]{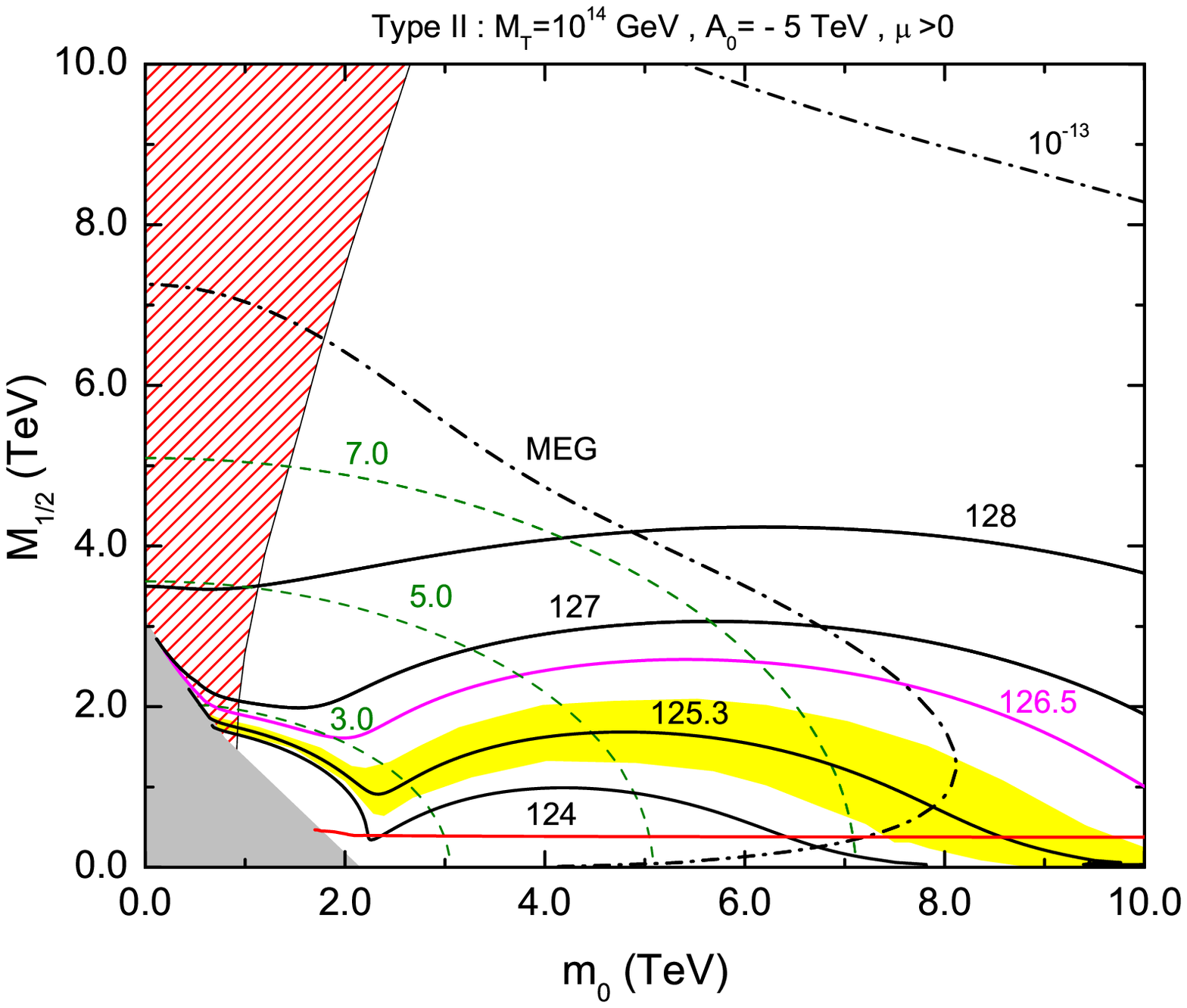}
\end{tabular}
\caption{\label{fig:m0m12typeII}Squark mass, Higgs mass
and Br($\mu\to e \gamma$) contours in the plane ($m_0,M_{1/2}$) for
CMSSM plus seesaw type II for two values of the seesaw scale $M_{T}$:
$M_{T}=10^{9}~\GeV$ (top) and $M_{T}=10^{14}~\GeV$ (bottom); as well
as two values of $A_0$: $A_0=0\,\TeV$ (left) and $A_0=-5~\TeV$ (right).
The values of Br($\mu\to e \gamma$) are orders of magnitude below the
expected experimental sensitivity in case of $M_{T}=10^{9}~\GeV$ and,
thus, are not shown. Note the change in scale compared to
Fig.~\ref{fig:m0m12typeI} (for further discussion see text).}
\end{figure}

\begin{figure}[t]
\begin{tabular}{ll}
\includegraphics[width=73mm]{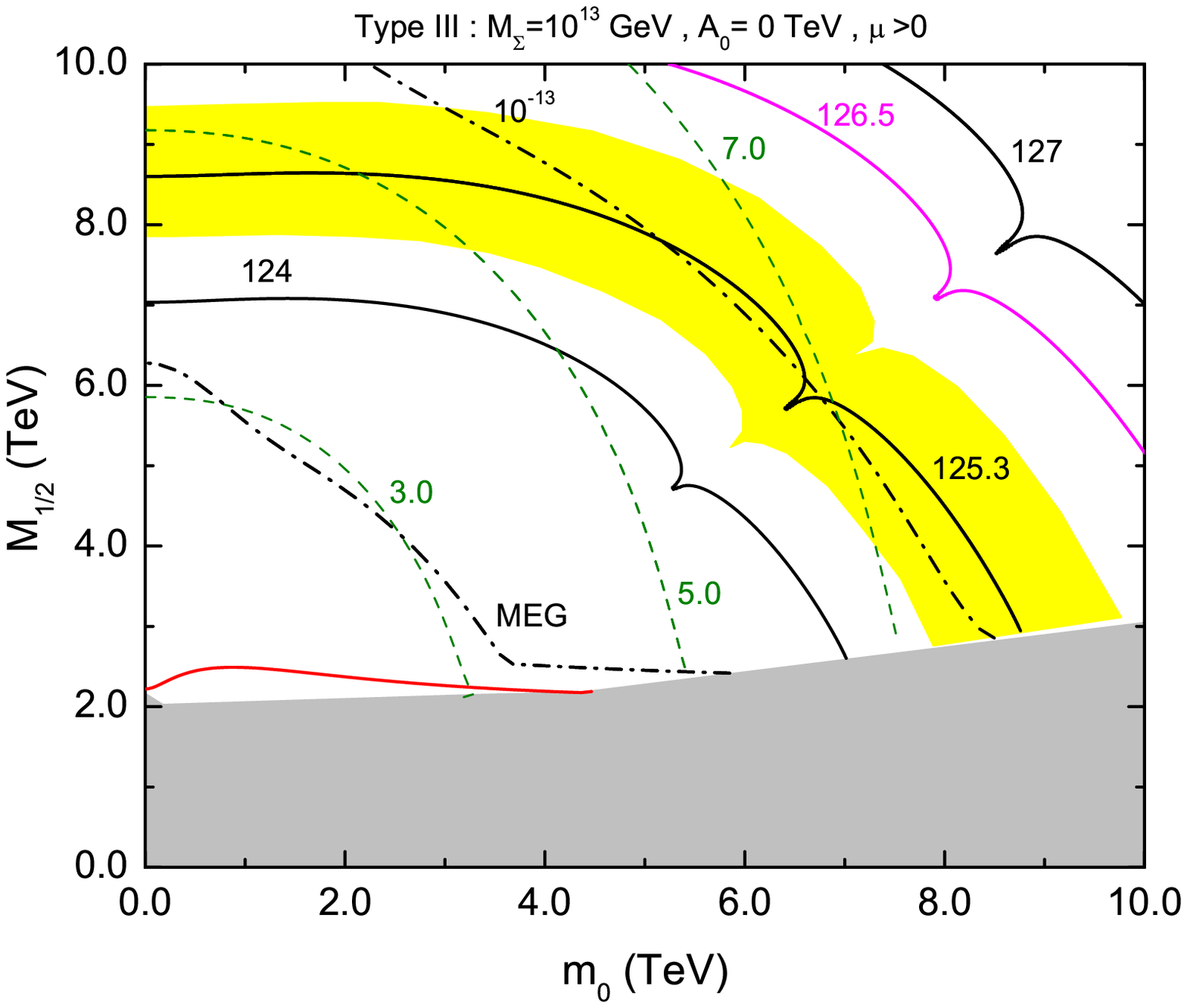} &
\hspace{-0.2cm}\includegraphics[width=73mm]{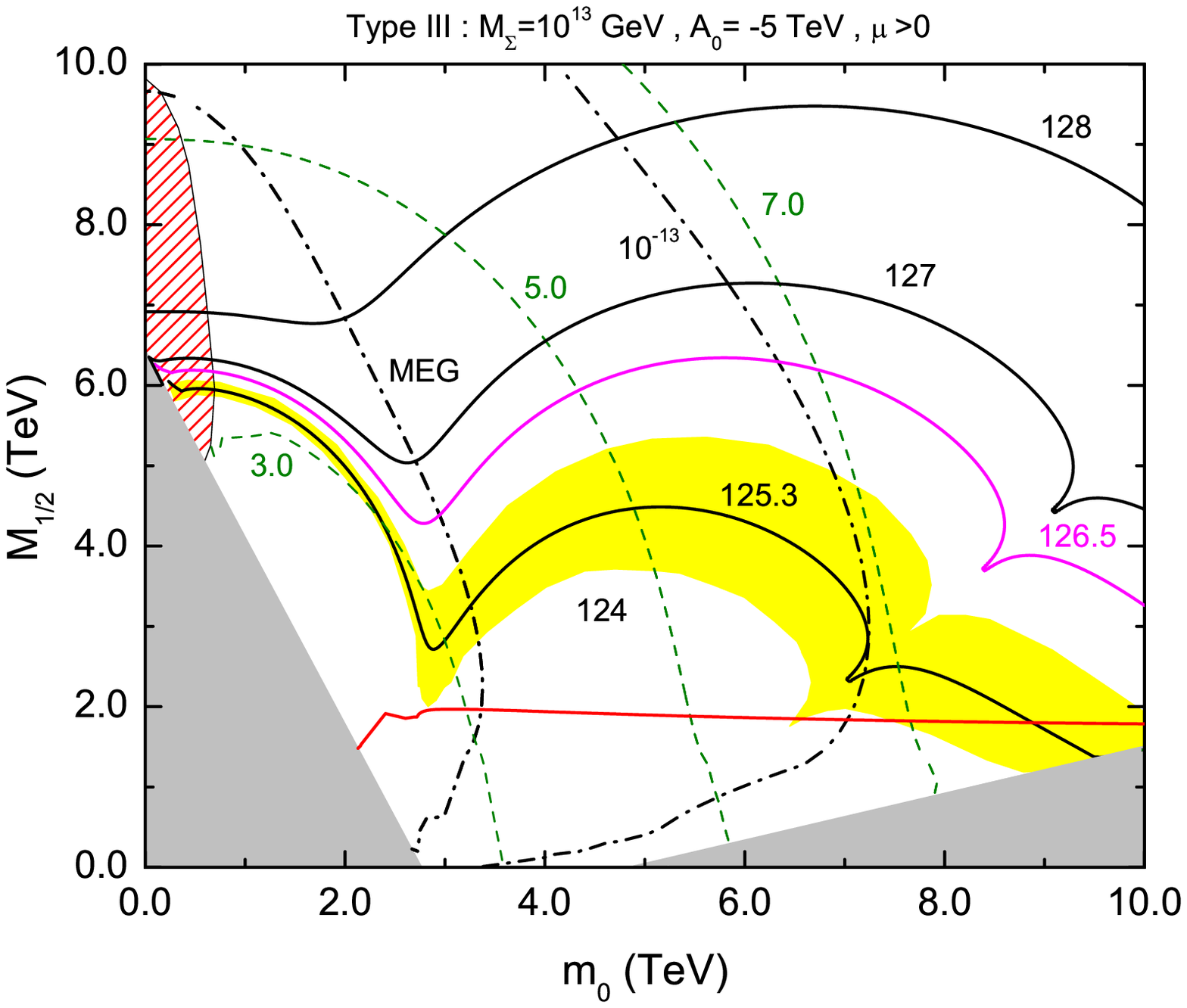}\\
\includegraphics[width=73mm]{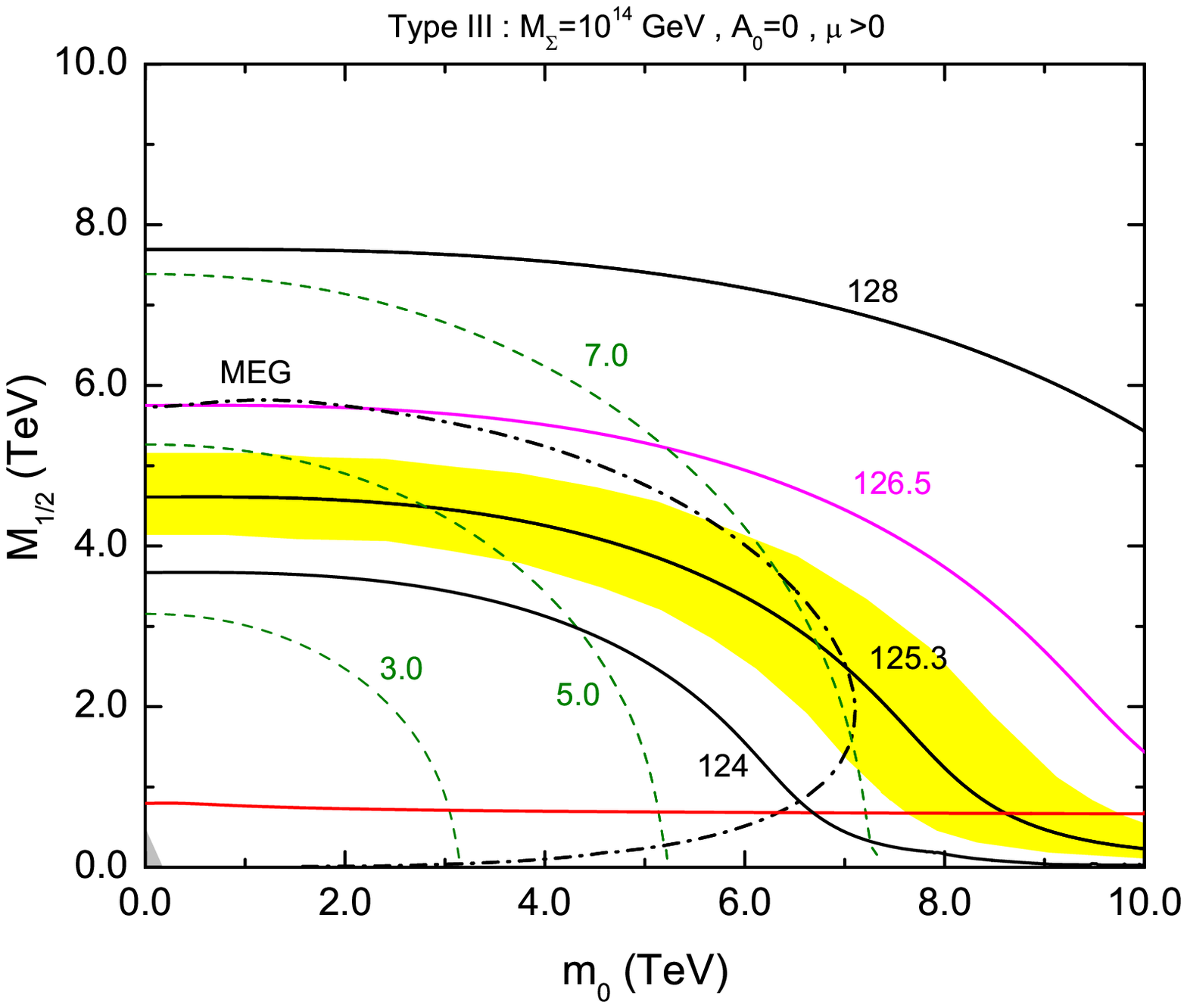}&
\hspace{-0.2cm}\includegraphics[width=73mm]{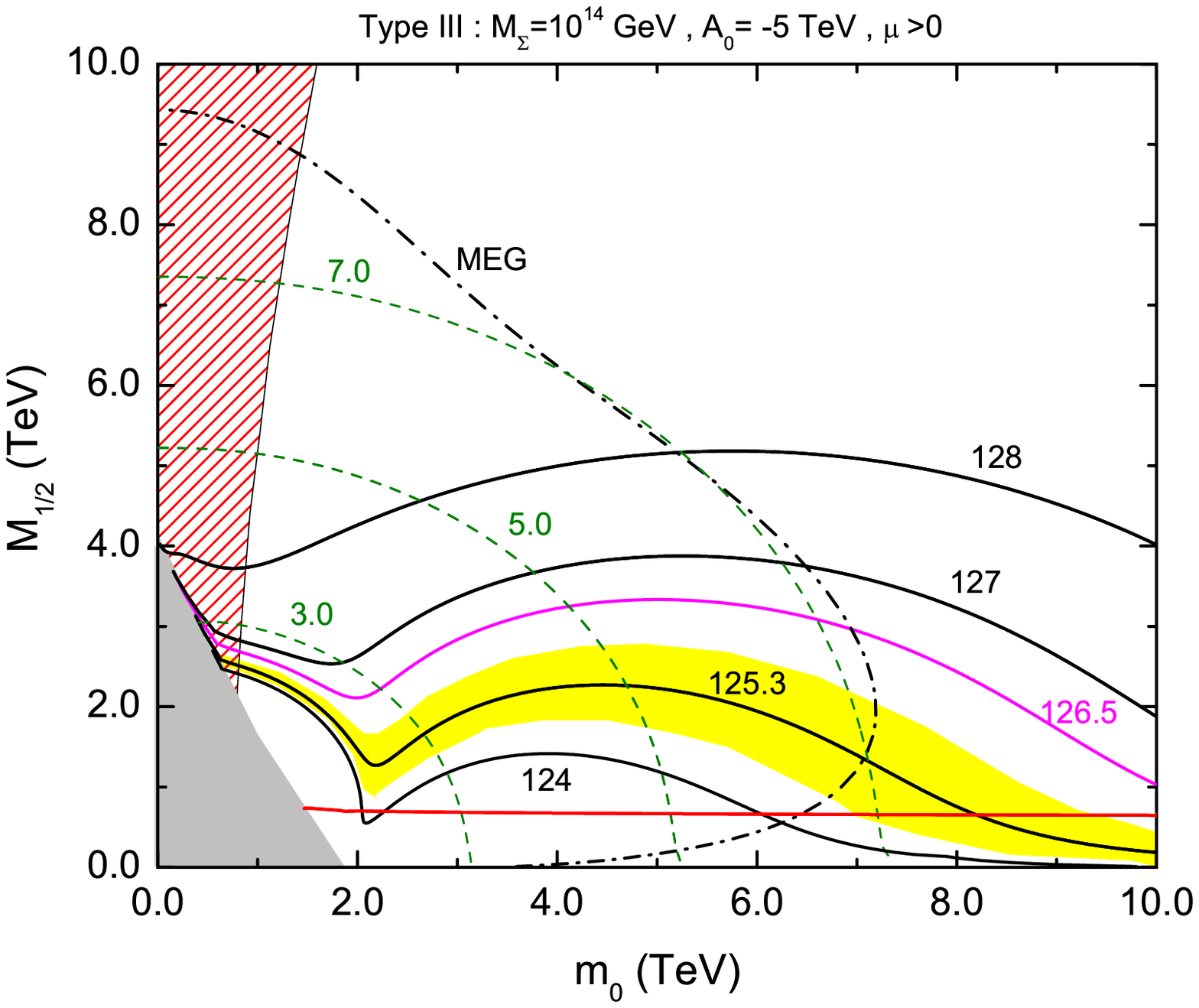}
\end{tabular}
\caption{\label{fig:m0m12typeIII}Examples of squark mass, Higgs mass
and Br($\mu\to e \gamma$) contours in the plane ($m_0,M_{1/2}$) for
CMSSM plus seesaw type III for two values of the seesaw scale $M_{SS}$:
$M_{\Sigma}=10^{13}~\GeV$ (top) and $M_{SS}=10^{14}~\GeV$ (bottom); as well
as two values of $A_0$: $A_0=0\,\TeV$ (left) and $A_0=-5~\TeV$ (right).
For a discussion see text.}
\end{figure}

In Fig.~\ref{fig:m0m12typeI} we show examples of squark mass, Higgs
mass and Br($\mu\to e \gamma$) contours in the plane ($m_0,M_{1/2}$)
for CMSSM plus seesaw type I, taking two extreme values of the seesaw
scale $M_{R}$, namely $10^{9}~\GeV$ (top) and $10^{14}~\GeV$ (bottom);
as well as two values of $A_0$: $A_0=0\,\TeV$ (left) and $A_0=-5~\TeV$
(right). Here, and in the corresponding figures for type-II
and type-III seesaw (Figs.~\ref{fig:m0m12typeII} and \ref{fig:m0m12typeIII}, respectively), we show contours of $m_{h^0}$ in the range $124-128$ GeV, which corresponds very roughly to the theoretical allowed range for a
calculated $m_{h^0} = 126$ GeV.
The hatched regions on the left lead to a charged LSP and, thus, are
not acceptable due to cosmological constraints (charged dark
matter). In the grey regions, EWSB is not possible in a consistent
way.  The solid lines show contours of $m_{h^0}$ at $124$, $125.3$
(central CMS value), $126.5$ (central ATLAS value) and 127 GeV, to
reflect the currently favoured region of $m_{h^0}$. The green dashed
lines correspond to constant average squark masses, defined as
\begin{equation}
\label{eq:defsq}
m_{\tilde q} = \frac{m_{\tilde d_L}+ m_{\tilde{d}_R}}{2}\,,
\end{equation}
while the (black) dash-dotted lines refer to the contours
$\text{Br}(\mu\to e \gamma)=2.4\times 10^{-12}$~(MEG) and $10^{-13}$.
The yellow region corresponds to values of $m_{h^0}$ in the CMS
interval $125.3\pm 0.6~\GeV$. Below the red solid line $m_{\tilde g}<800~\GeV$.

For $m_{h^0}=125~\GeV$ we find squark masses in the range of typically
$5~\TeV$ for $A_0=0\,\TeV$ and as low as $2~\TeV$ for
$A_0=-5~\TeV$. However, considerably larger squark masses, ${\cal
  O}(10)~\TeV$, can be found in the CMS preferred window of Higgs
mass.  This is consistent with the findings of previous works on the
CMSSM~\cite{Arbey:2011ab,Baer:2011ab,Ellis:2012aa} and in agreement
with expectations. If this scenario is indeed realized in nature one
expects to observe squarks at the LHC with $\sqrt{s}=14~\TeV$ (LHC14)
only for the largest values of $A_0$. Still, even for $A_0=-5~\TeV$
large parts of the allowed parameter space in squark and gluino masses
will remain unexplored by LHC14.

Fig.~\ref{fig:m0m12typeI} shows also contours of ${\rm Br}(\mu\to e \gamma)$
assuming (a) a degenerate RH neutrino spectrum with $\bR={\bf 1}$
(see Section~\ref{sec:modelI}) and (b) low-energy neutrinos fitted
with a normal hierarchy spectrum and mixing angles within the allowed
range~\cite{Tortola:2012te}. It is well-known that different choices of
$\theta_{13}$ can lead to values of ${\rm Br}(\mu\to e \gamma)$ differing by a considerable
factor~\cite{Antusch:2006vw,Calibbi:2006ne}. However, we fix
$\theta_{13}$ according to the results of~\cite{Tortola:2012te}, where
a global fit to all available experimental data gives a best-fit value
of $\sin^2 \theta_{13} = 0.026$ in case of a neutrino spectrum with
normal hierarchy. We note that with degenerate
RH neutrinos and $\bR={\bf 1}$, a complete cancellation of
Br($\mu\to e \gamma$) is no longer possible within the 3$\sigma$
allowed range of $\sin^2\theta_{13}$ \cite{Tortola:2012te}. In case of
$M_{R}=10^{9}~\GeV$, Br($\mu\to e \gamma$) is orders of magnitude
below the expected experimental sensitivity and, thus, contours are
not shown. However, if $M_{R}=10^{14}~\GeV$, Br($\mu\to e \gamma$) is
well within the current expected sensitivity of MEG. Therefore, for
the CMSSM with a seesaw type I and $m_{h^0}=125$, MEG already provides
an upper limit on $M_{R}$ of the order of $10^{14}~\GeV$,
despite the fact that sleptons in the CMSSM are relatively heavy in
the allowed parameter space. We notice that the constraints from
Br($\mu\to e \gamma$) are, in general, more stringent for large values
of $A_0$.

In Fig.~\ref{fig:m0m12typeII} we show the results in the plane
($m_0,M_{1/2}$) for type II seesaw with $M_T=10^9,10^{14}~\GeV$ and
$A_0=0,-5~\TeV$. When $M_{T}=10^{14}~\GeV$, the results for $m_{h^0}$
and the squark masses are very similar to the CMSSM ones, although
some small shifts are visible upon closer inspection (see also below).
On the other hand, the contours for Br($\mu\to e \gamma$) are
different from those in Fig.~\ref{fig:m0m12typeI}. This is in
agreement with expectations~\cite{Hirsch:2008dy,Hirsch:2008gh}, since
in type I neutrino masses scale as the square of the Yukawa couplings
whereas in type II neutrino masses are linearly proportional to
$\bY_T$ [see Eqs.~(\ref{eq:mnuI}) and (\ref{eq:ssII})], while the RG
running of the LFV soft masses depends quadratically on the Yukawas in
both cases. Note, that in Fig.~\ref{fig:m0m12typeII} we have used
$\lambda_{2}=0.5$. A value of $\lambda_2=1$ would lead to smaller
values of Br($\mu\to e \gamma$) by (roughly) a factor of four. Much
larger values of $\lambda_2$ are not allowed, if the theory is to
remain perturbative up to the GUT scale.

\begin{figure}[t]
\begin{tabular}{ll}
\includegraphics[width=73mm]{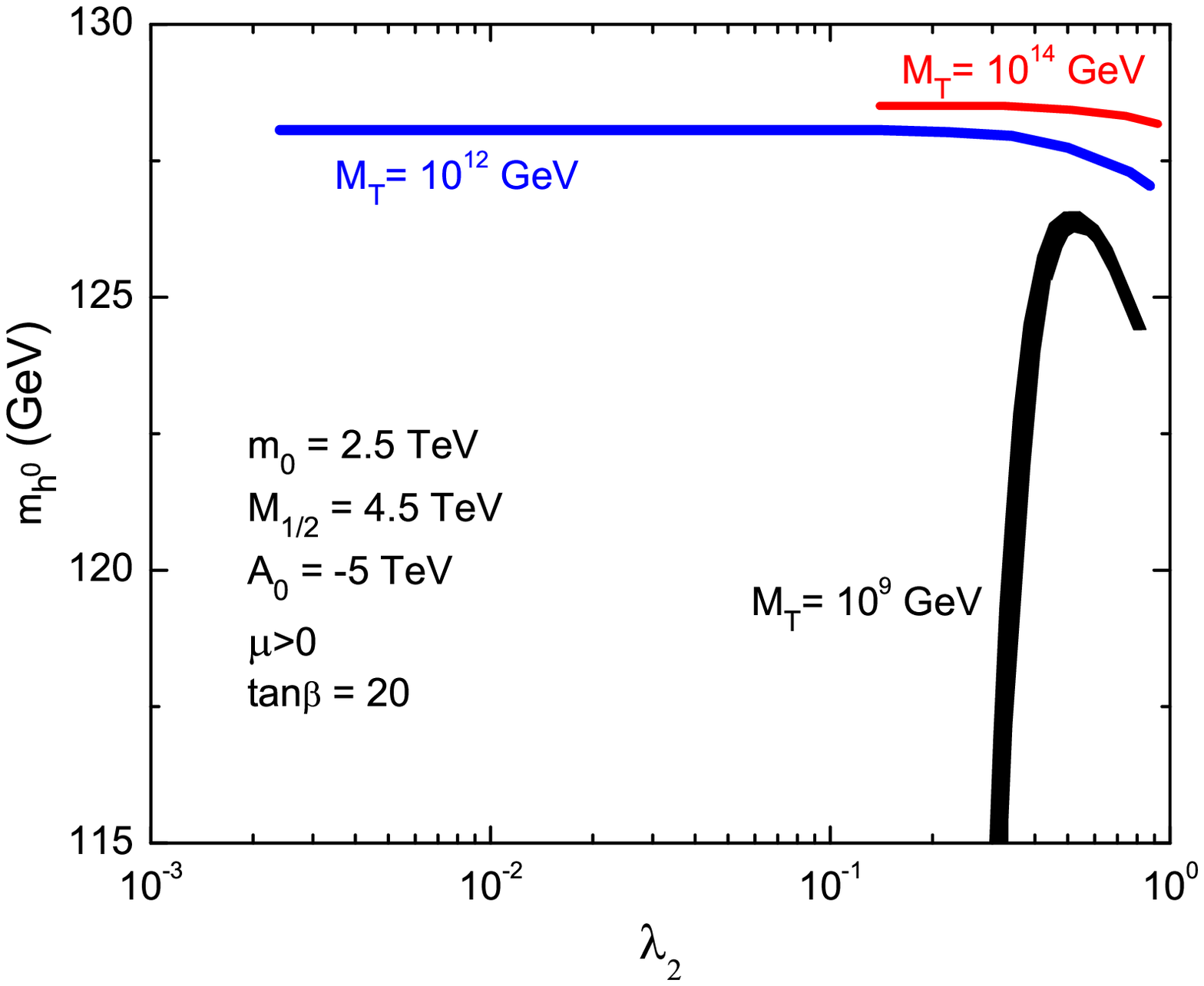} &
\hspace*{-0.2cm}\includegraphics[width=73mm]{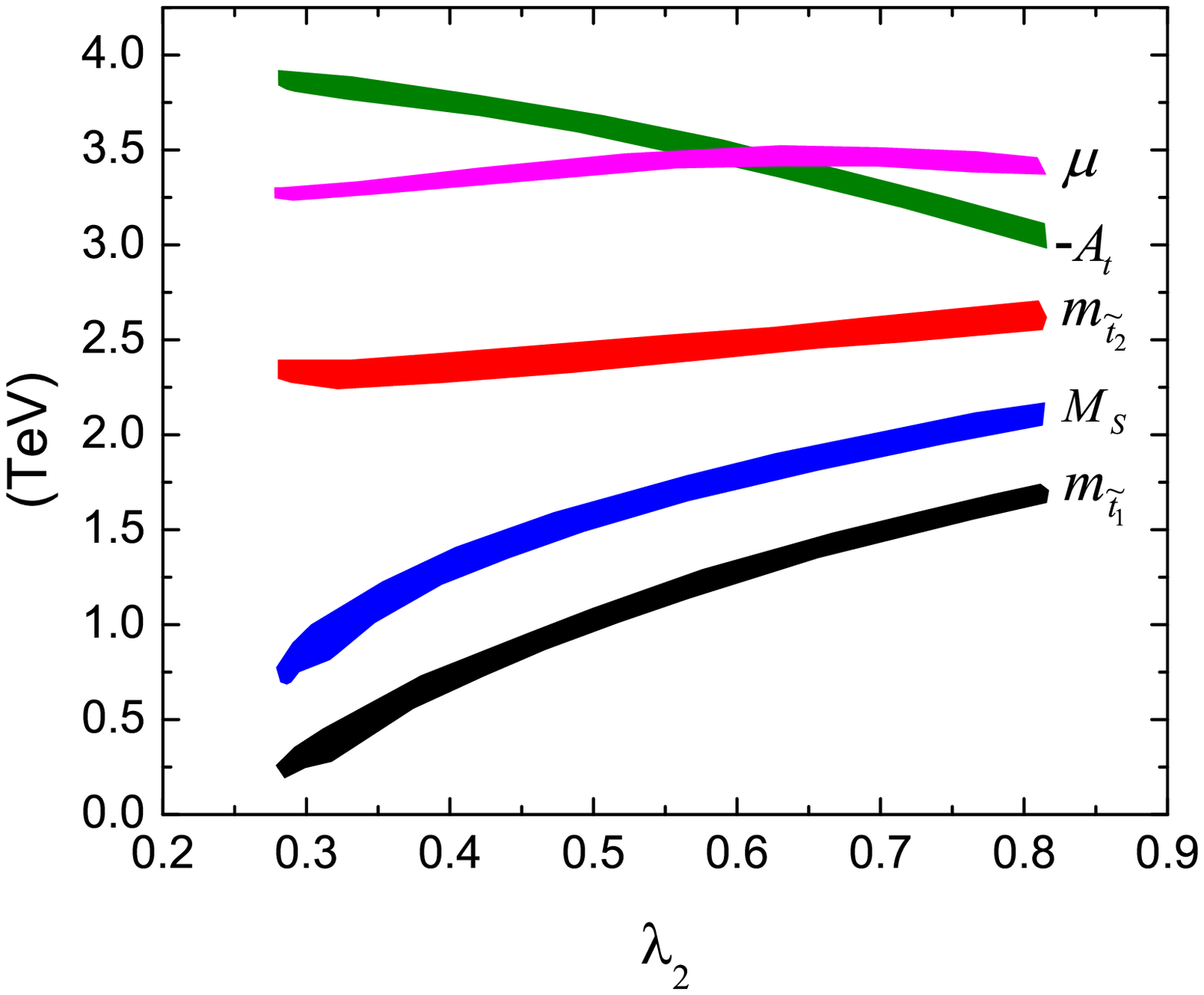} \\
\includegraphics[width=73mm]{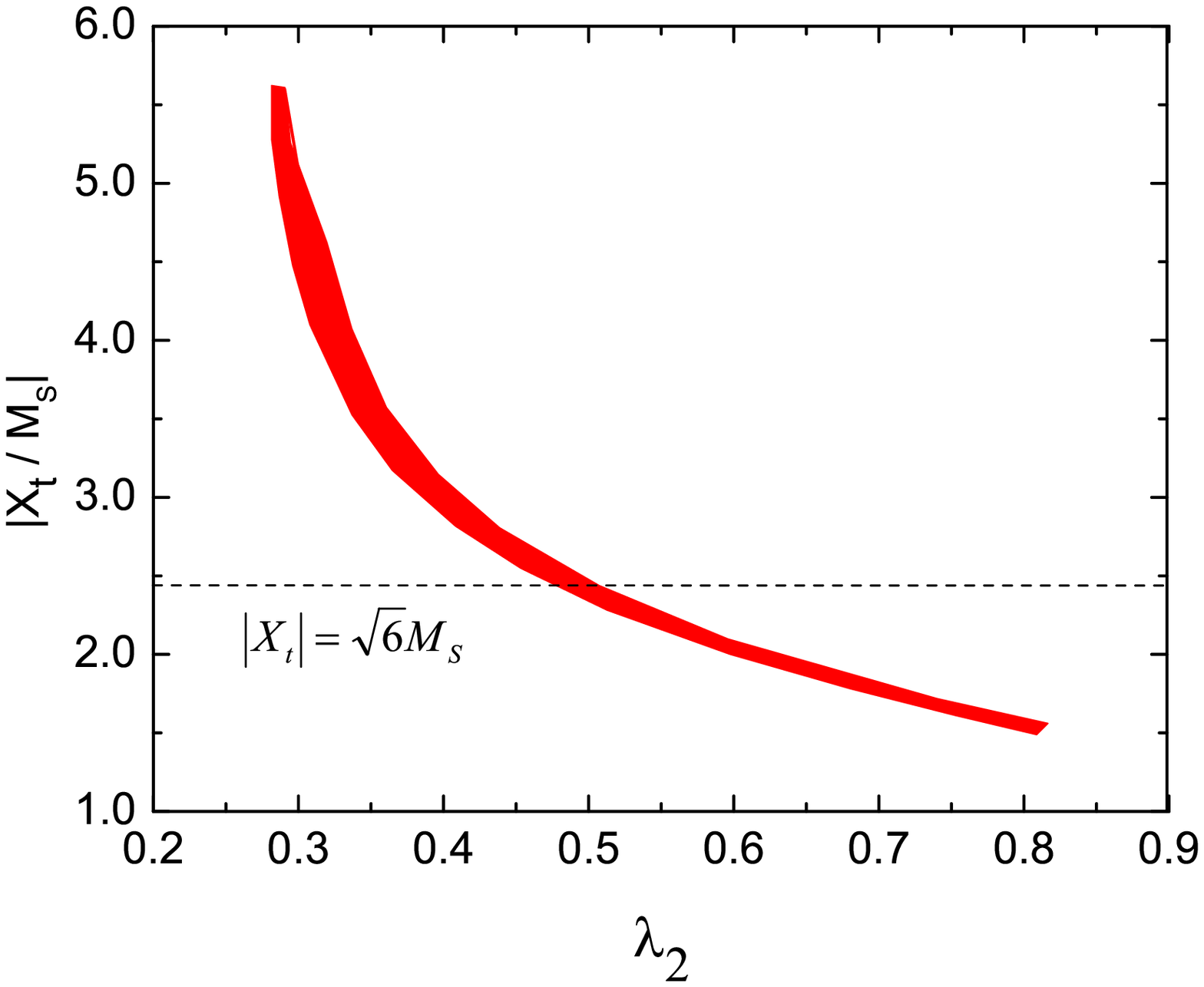} &
\hspace*{-0.2cm}\includegraphics[width=73mm]{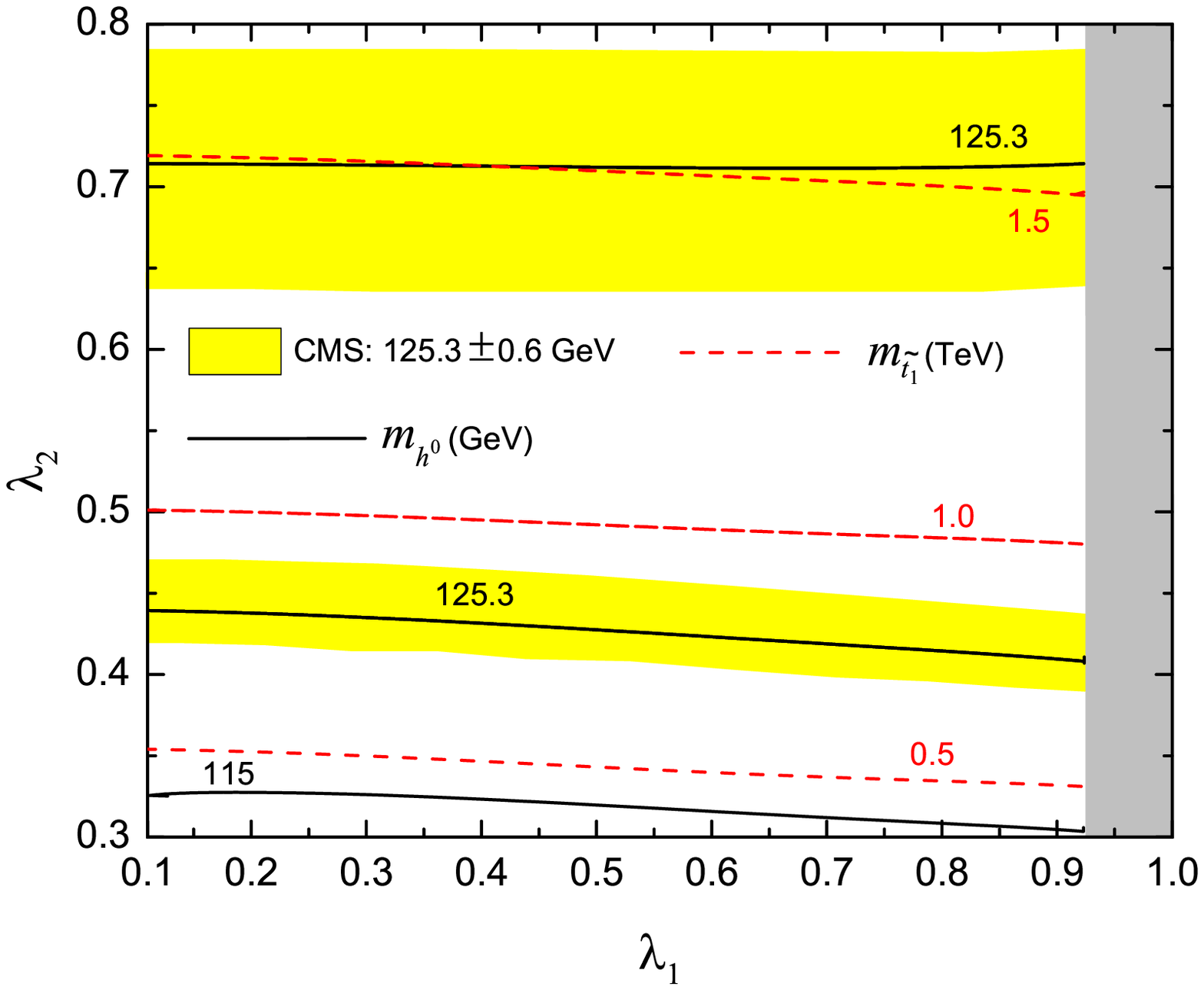}
\end{tabular}
\caption{\label{fig:l1l2TypeII}Dependence of several low-energy SUSY
  parameters on the couplings $\lambda_{1,2}$ for the type II seesaw
  scenario. The results are shown for a specific point in the SUSY
  parameter space with $m_0=2.5\,{\rm TeV}$, $M_{1/2}=4.5\,{\rm TeV}$,
  $A_0=-5\,{\rm TeV}$, $\tan\beta=20$ and $\mu>0$. Top left: Higgs
  mass as a function of $\lambda_2$ for $M_T=10^9\,{\rm GeV}$ (black),
  $M_T=10^{12}\,{\rm GeV}$ (blue) and $M_T=10^{14}\,{\rm GeV}$
  (red). Top right: $\lambda_2$ dependence of the stop masses
  $m_{\tilde{t}_{1,2}}$ (and their geometric average $M_S$), the
  Higssino mass parameter $\mu$ and the top-trilinear term
  $A_t$. Bottom left: stop mixing parameter $X_t$ as function of
  $\lambda_2$. Bottom right: contours of the Higgs (black solid) and
  lightest stop (red dashed) masses in the $(\lambda_1,\lambda_2)$
  plane for $M_T=10^9\,{\rm GeV}$. The yellow regions corresponds to
  the CMS Higgs mass interval $m_{h^0}=125.3\pm 0.6\,{\rm GeV}$. There are
  two CMS allowed contours in the lower right plot, since $m_{h^0}$
  first increases then decreases with $\lambda_2$ for $M_T=10^9\,{\rm
    GeV}$, compare to the figure in the upper left.}
\end{figure}

For $M_{T}=10^{9}~\GeV$, on the other hand, the results look
drastically different. All Higgs (and squark) mass contours are
shifted to larger values of $m_0$ and $M_{1/2}$. This is in agreement
with the previous observation that lower values of the seesaw
scale lead to lighter sparticles for the same point in CMSSM parameter
space (see discussion of Fig.~\ref{fig:exa}).  As a consequence, the
Higgs becomes lighter. To compensate for this downward shift in the
SUSY spectrum one has to increase the parameters $m_0$ and/or
$M_{1/2}$. However, while a low type II scale now requires very large
$m_0$ and/or $M_{1/2}$, the resulting squark (and gluino) contours in
the interesting range of $m_{h^0}$ are similar to those found in the
CMSSM.  This stems from the fact that $m_{h^0}=125~\GeV$ requires
again squark masses in the range of (at least) $5~\TeV$ for $A_0=0\,\TeV$
and $2~\TeV$ for $A_0=-5~\TeV$. This is not surprising since the Higgs
mass is sensitive only to physical masses and mixings. However, as we
will discuss below, there are some potentially interesting differences
in the spectra due to the different RG running in the CMSSM and the
SUSY type II seesaw.

For the lowest value of $M_{T}$, where the spectrum distortions are
larger, Br($\mu\to e \gamma$) is again negligible. Thus, an upper
limit on Br($\mu\to e \gamma$) provides an upper limit on $M_{T}$ for
any given value of the Higgs mass. A measurement of Br($\mu\to e
\gamma$) fixes a combination of $\lambda_2$ and $M_{T}$ for fixed
$m_{h^0}$. On the other hand, a lower limit on $m_{h^0}$ provides a
lower limit on a combination of $m_0$, $M_{1/2}$ and $A_0$ for any
fixed choice of $M_{T}$. Note that, contrarily to what happens in type
I and III, in type II seesaw (with a single $15$-plet pair) low-energy
neutrino parameters essentially determine $\bY_T$ in a way that large
cancellations in the LFV soft masses are not possible. Moreover, when
LFV in the soft masses is generated by $\bY_T$ only, the large value
of $\sin\theta_{13}$ provided by the latest global analysis of
neutrino oscillation data together with the present MEG bound on
$\mu\to e\gamma$ set an upper limit on the radiative $\tau$ decays
$\tau\to \mu (e)\gamma$ which is out of the reach of future
experiments~\cite{Rossi:2002zb,Joaquim:2006mn,Joaquim:2006uz,Hirsch:2008gh,Joaquim:2009vp,Brignole:2010nh}.

In the SUSY type II seesaw, the heavy triplet states $T$ and $\bar{T}$
couple to the MSSM Higgs sector through the superpotential couplings $\lambda_{1,2}$ [see
Eq.~(\ref{eq:broken})]. We therefore expect these parameters to affect
the Higgs mass to some extent. Obviously, since $T$ and $\bar{T}$ are
very heavy, the effect of $\lambda_{1,2}$ on the low-energy SUSY
masses is indirect and originates from RG corrections induced on the
SUSY parameters between $M_{GUT}$ and $M_T$. Consequently, these
corrections are typically larger for smaller $M_T$. In
Fig.~\ref{fig:l1l2TypeII} we show the dependence of several parameters
relevant for the computation of $m_{h^0}$ as a function of $\lambda_1$
and $\lambda_2$ (taken at the scale $M_T$), for a specific point of the SUSY parameter space (see
caption). In the top-left panel we show a plot of $m_{h^0}$ versus
$\lambda_2$ (and varying $\lambda_1$ from 0.1 to the maximum allowed
by perturbativity) for $M_T=10^9,~10^{12},~10^{14}~\GeV$. As expected,
the impact of $\lambda_2$ on the Higgs mass is only significant for
the case with $M_T=10^9~\GeV$. In the remaining two examples, a mild
dependence on $\lambda_2$ is observed when the value of this parameter
is very close to the Landau pole. From this plot one can also conclude
that the effect of $\lambda_1$ on $m_{h^0}$ is small, since the
thickness of the lines (which reflects the variation of $m_{h^0}$ on
$\lambda_1$) is not too pronounced. In view of this, we
will only comment on the $\lambda_2$-dependence of $m_{h^0}$ for $M_T=10^9~\GeV$.

The top-right panel of Fig.~\ref{fig:l1l2TypeII} shows the
variation of some relevant parameters with $\lambda_2$ (see caption for more details). We first note
that while $\mu$ and $m_{\tilde t_{1,2}}$ (and consequently $M_S$)
increase with increasing $\lambda_2$, $|A_t|$ decreases (here $A_t$ is
always negative since $A_0<0$). This behaviour can be qualitatively
understood by looking at the type II seesaw RGEs for the soft masses
and trilinear terms. In particular, we notice that the RGE for $A_t$
contains a term proportional to $|\lambda_2|^2A_t$ which, at leading-log approximation, induces a
correction to the top trilinear given by
\begin{equation}
\label{eq:Atapp}
\Delta A_t=-\frac{3\,y_t |\lambda_{2}|^2}{8\pi^2}
A_0\ln\left(\frac{M_{GUT}}{M_T}\right)\,,
\end{equation}
which is positive for $A_0<0$. This explains why $-A_t$ decreases with
$\lambda_2$ and, consequently, why $X_t$ decreases~\footnote{Notice
that in this case $X_t\sim -A_t$ since $\mu\sim -A_t$ and $\cot\beta=0.05$.}. The
behaviour of $\mu$ can be traced taking into account that the
Higgs soft masses $m^2_{H_{d,u}}$ receive a contribution which amounts
to:
\begin{equation}
\label{eq:m2Happ}
\Delta m^2_{H_{d,u}}=-\frac{9m_0^2+3A_0^2}{8\pi^2}
|\lambda_{1,2}|^2\left(\frac{M_{GUT}}{M_T}\right)\,.
\end{equation}
Notice that, if not too small, the parameter $\lambda_1$ can act on
$m^2_{H_d}$ as the top Yukawa coupling does on $m^2_{H_u}$ bringing it
to negative values at low-energies. In fact, we observe that for
$\lambda_1$ large, $m^2_{H_d}$ is also negative at the EW scale. We
recall from the EWSB symmetry breaking condition:
\begin{equation}
\label{eq:muEWSB}
\mu^2=\frac{m^2_{H_d}-\tan^2\beta\,m^2_{H_u}}{\tan^2\beta-1}-\frac{m_Z^2}{2}.
\end{equation}
As $\lambda_2$ increases, $m^2_{H_u}$ becomes more negative and
$m^2_{H_d}$ decreases, going from positive to negative values. This
leads to an increasing of the value of $\mu$ with
$\lambda_2$.

Although not affected directly by $\lambda_2$, the stop
masses $m_{\tilde{t}_{1,2}}$ (and, thus, the dynamical scale $M_S$)
increases with that parameter mainly due to positive RG corrections in
$(\bmm_{\tilde{Q},\tilde{u}^c})_{33}$. The results for $|X_t/M_S|$ as function of
$\lambda_2$ are shown in the bottom-left panel of
Fig.~\ref{fig:l1l2TypeII}. Together with Eq.~(\ref{eq:mhapp}), these
results allow us to understand the behaviour of $m_{h^0}$ with
$\lambda_2$ shown in the top-left panel. In particular, we stress that
for $\lambda_2\simeq 0.5$ we have $|X_t|=\sqrt{6}M_S$, which
corresponds to the ``maximal mixing'' scenario with maximised
$m_{h^0}$ (see the top-left panel). Finally, in the bottom-right
panel, the contours of $m_{h^0}$ and $m_{\tilde{t}_1}$ are shown in
the $\lambda_{1,2}$ plane for $M_T=10^9~\GeV$. The results confirm
that while both $m_{h^0}$ and $m_{\tilde{t}_1}$ depend reasonably
strong on $\lambda_2$, their dependence on $\lambda_1$ is almost
negligible. In particular, the effect of $\lambda_2$ on $m_{h^0}$ can
be much larger than its theoretical uncertainty.

In Fig.~\ref{fig:m0m12typeIII} we show the results in the
($m_0,M_{1/2}$) plane for type III seesaw with
$M_{\Sigma}=10^{13},\,10^{14}~\GeV$. For lower values of the 24-plet
mass no solutions consistent with perturbativity exist.  Since in type
III all SUSY masses run strongly towards smaller values when
$M_{\Sigma}$ is lowered, already for $M_{\Sigma}=10^{13}~\GeV$ the
spectrum distortions with respect to the type I case are as large (or
larger) as those found for type II with $M_{T}=10^{9}~\GeV$ (compare
Figs.~\ref{fig:m0m12typeII} and \ref{fig:m0m12typeIII}).  As in type I
and II, multi-TeV squarks (and gluinos) are required to explain
$m_{h^0}\simeq 125~\GeV$. Still, depending on $M_{\Sigma}$, the
relations among sparticle masses are changed. It is interesting to
note that Br($\mu\to e \gamma$) provides a particularly strong
constraint for type III~\cite{Esteves:2009qr,Biggio:2010me}. In case of
$M_{\Sigma}=10^{14}~\GeV$ (bottom plots in
Fig.~\ref{fig:m0m12typeIII}), an improvement of Br($\mu\to e \gamma$)
to the level of $\simeq 10^{-13}$ (within the reach of MEG) would
exclude the type III seesaw with degenerate $24$-plets, $\bR={\bf 1}$
and a Higgs mass lying in the ATLAS and CMS range. In the particular
case of $A_0=-5~{\rm TeV}$ (bottom-right plot) most of the CMS
preferred region (in yellow) is already excluded by the constraint
from MEG, which also excludes squark masses below $\sim 7~\TeV$.

\begin{figure}[t]
\begin{tabular}{ll}
\includegraphics[width=73mm]{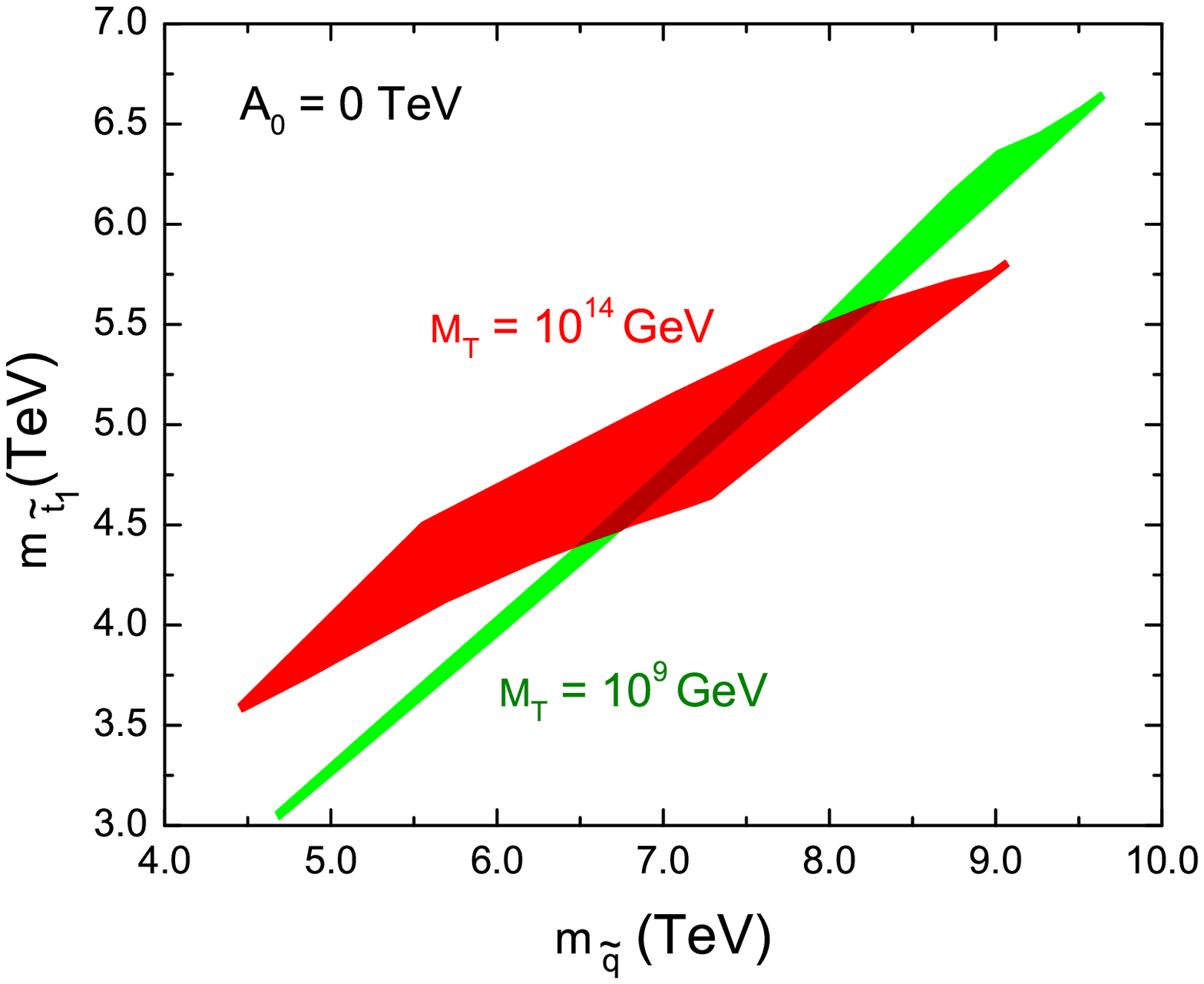} &
\hspace*{-0.2cm}\includegraphics[width=73mm]{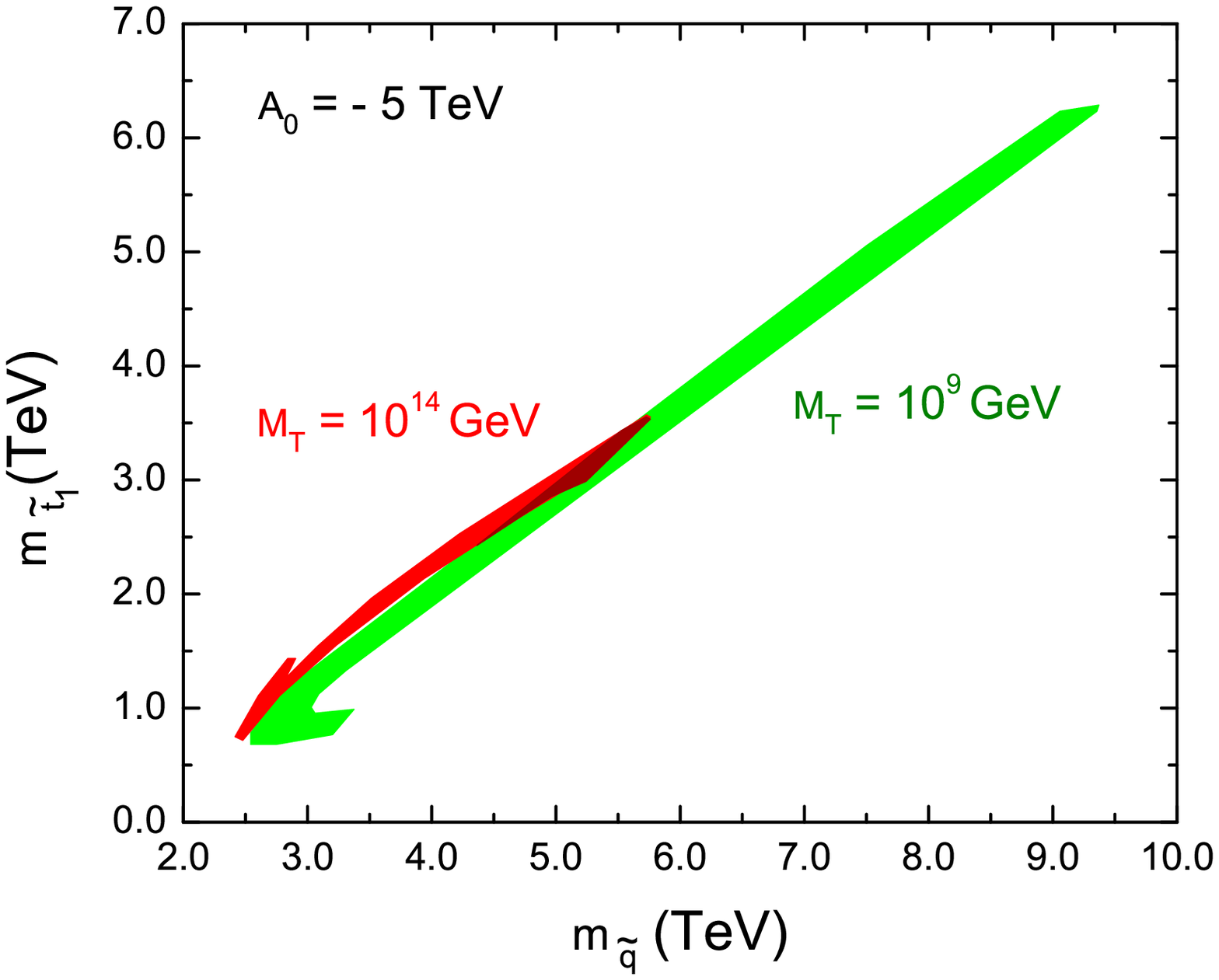}\\
\includegraphics[width=73mm]{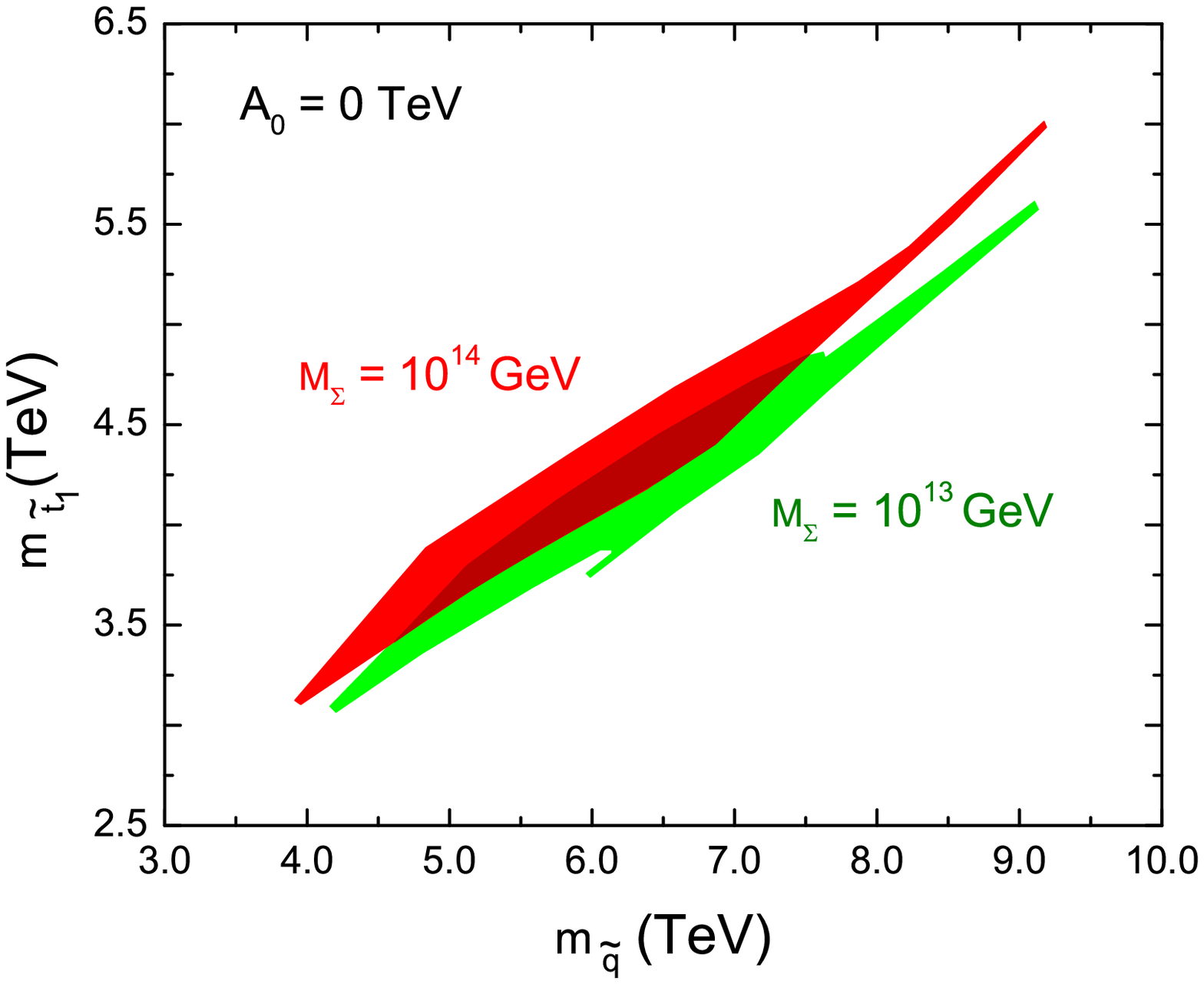}&
\hspace*{-0.2cm}\includegraphics[width=73mm]{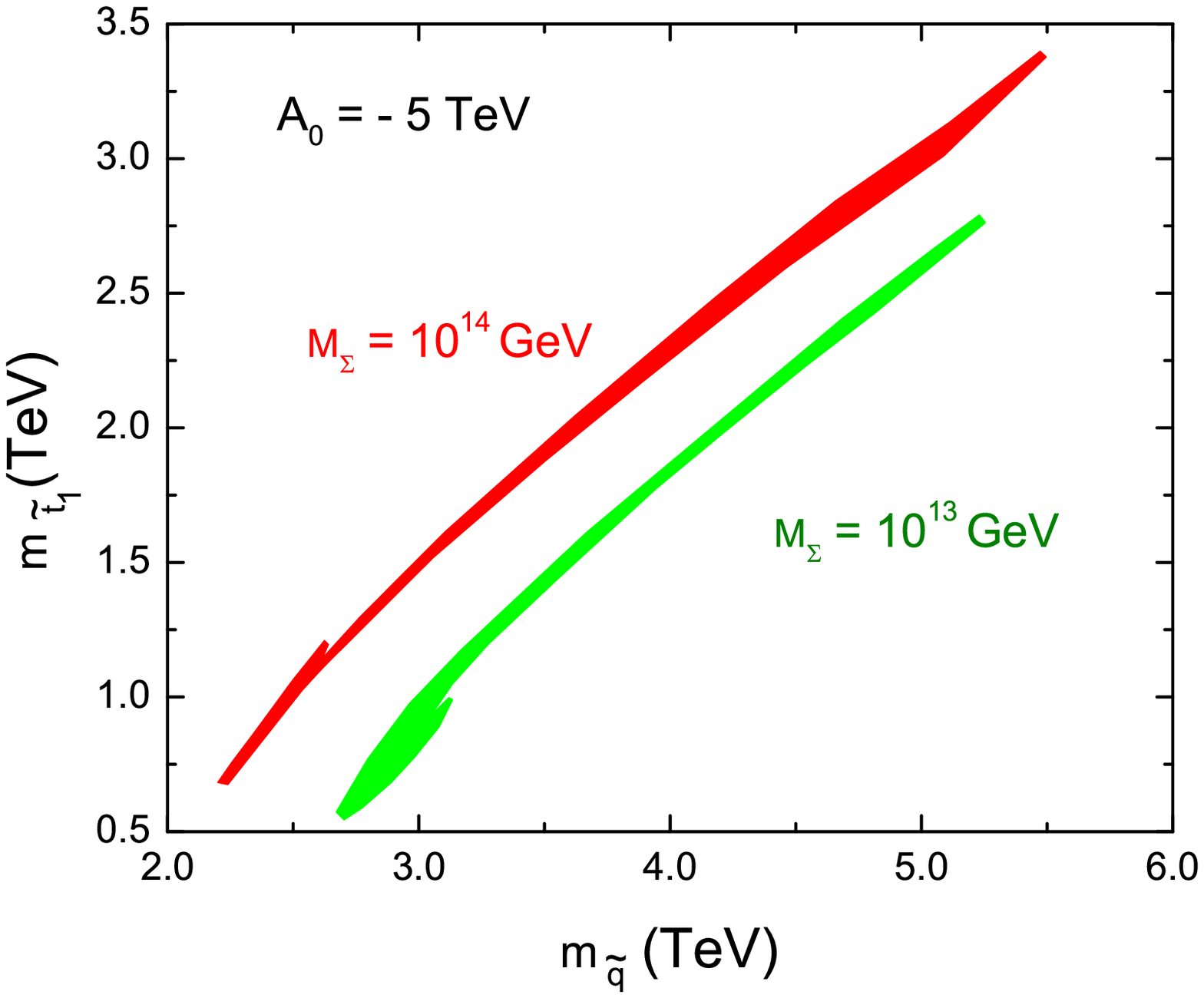}
\end{tabular}
\caption{\label{fig:mSqvmsToptypeIII} Allowed ranges of the lightest
stop versus squark mass compatible with a $125.3 \pm 0.6\,{\rm GeV}$
Higgs (CMS range) for type II (top) and type III (bottom) seesaws with
$A_0=0~\TeV$ (left) and $A_0=-5~\TeV$ (right). The red (green) regions
are for $M_{T,\Sigma}= 10^{14}$ ($M_{T}= 10^{9}~\GeV$ and $M_{\Sigma}=
10^{13}$) GeV.}
\end{figure}

\begin{figure}[t]
\begin{tabular}{ll}
\includegraphics[width=73mm]{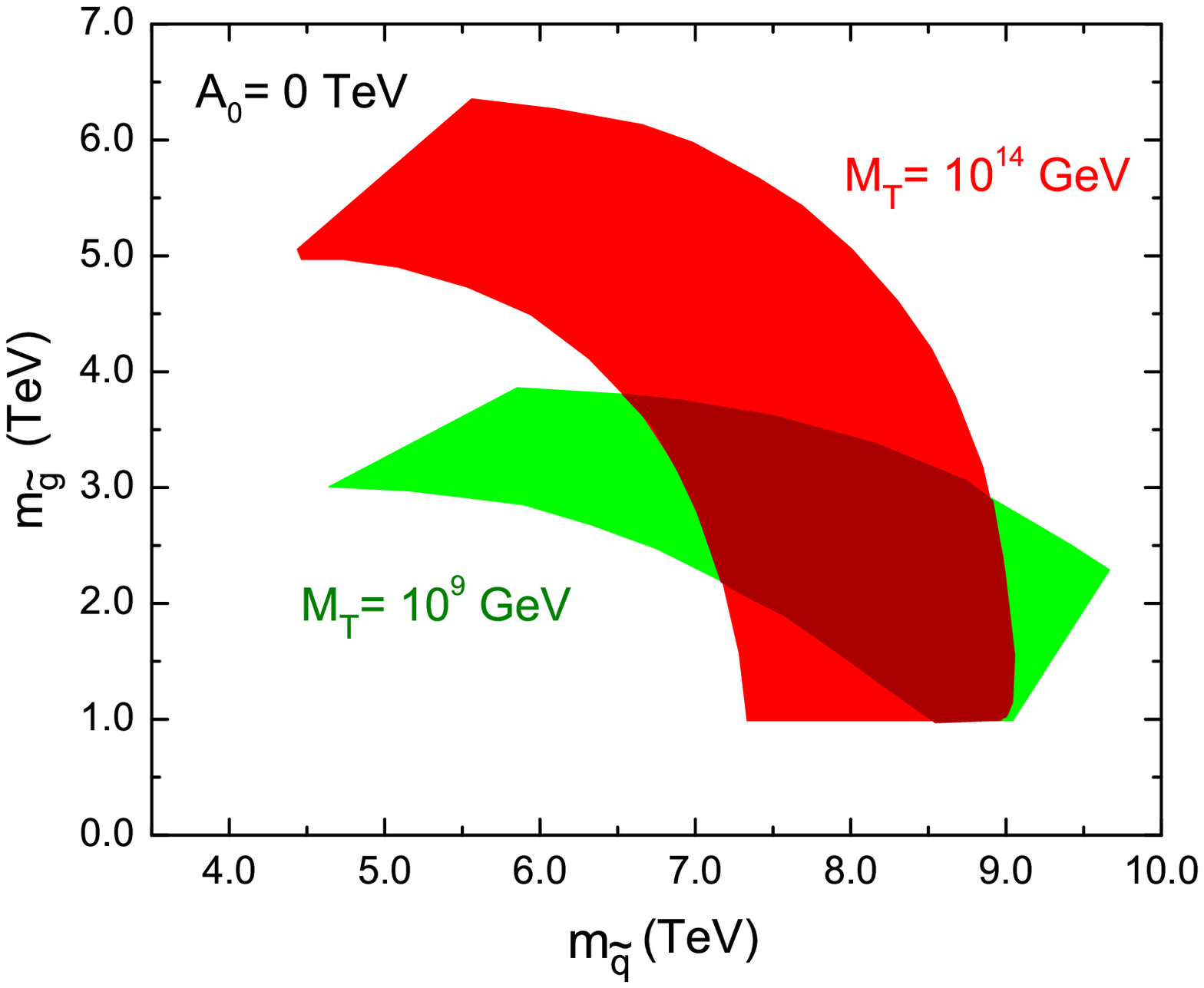} &
\hspace{-0.2cm}\includegraphics[width=73mm]{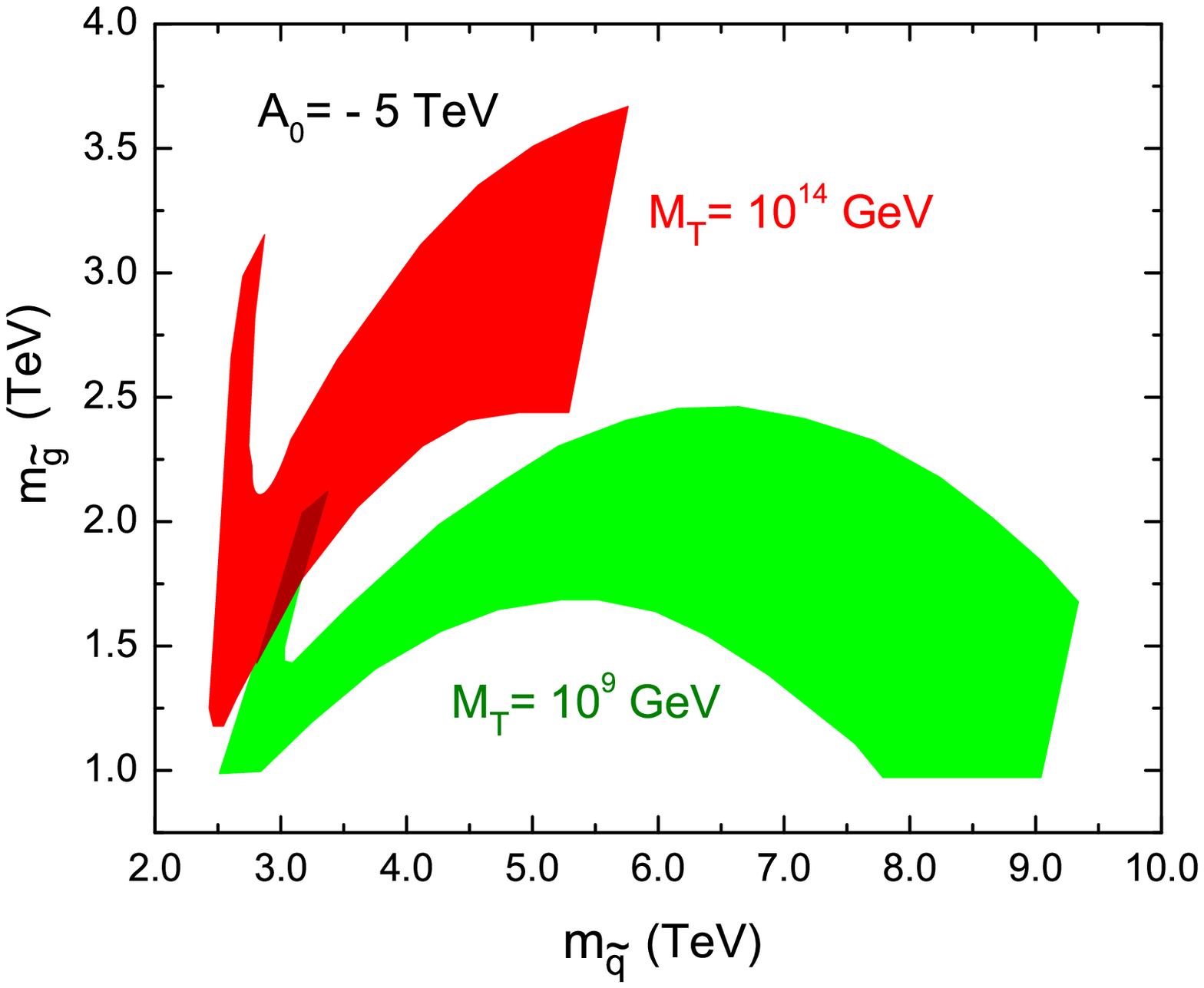}\\
\includegraphics[width=73mm]{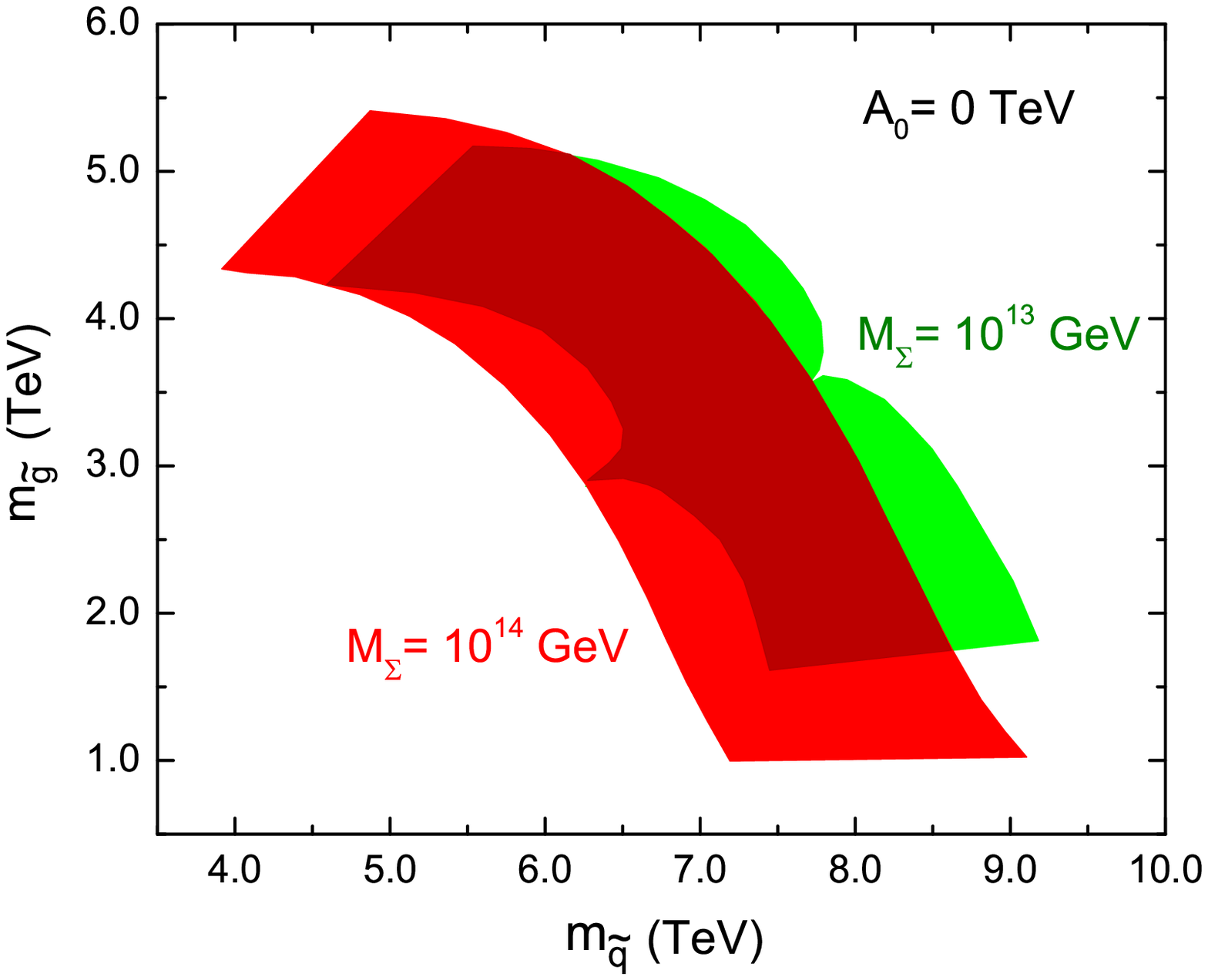} &
\hspace{-0.2cm}\includegraphics[width=73mm]{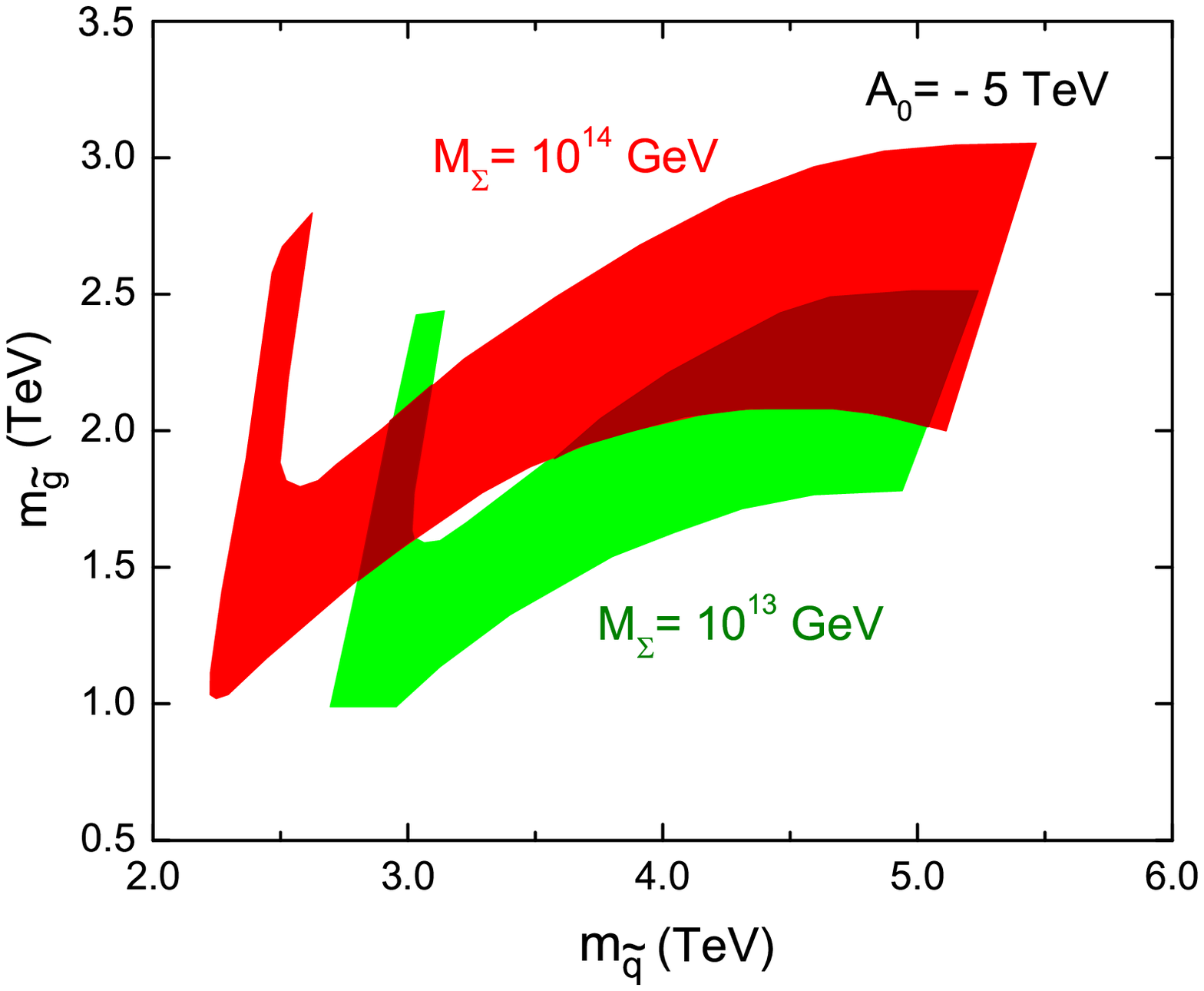}
\end{tabular}
\caption{\label{fig:msqmgII}Allowed regions in the $(m_{\tilde
    q},m_{\tilde g})$ plane for seesaw type II (top) and III
  (bottom). In the left (right) panels $A_0=0\,\TeV$
  ($A_0=-5\,\TeV$). We consider two extreme values of the seesaw scale
  in all cases as well as a fixed interval for the Higgs mass, namely
  $m_{h^0}=125.3 \pm 0.6$ GeV. For a discussion see text.}
\end{figure}

We now turn to a discussion on differences found in the physical
masses for the different seesaw setups. Since for type I the spectra
are practically the same as in the CMSSM (which has been discussed at
length in the literature) we focus on type II and III seesaws in the
following. In Fig.~\ref{fig:mSqvmsToptypeIII} we show the allowed
ranges for the lightest stop mass $m_{\tilde{t}_1}$ and the average
squark mass, defined in Eq.~(\ref{eq:defsq}), for seesaws of type II
and III with a Higgs mass in the CMS range $125.3 \pm 0.6~\GeV$ (the same range is considered in
Fig.~\ref{fig:msqmgII})\footnote{While this paper was in the review process, the ATLAS collaboration released the result $m_{h^0} = 126.0 \pm 0.4 \pm 0.4$ GeV~\cite{Aad:2012gk}. Using this Higgs mass range would lead to allowed regions similar to those shown in Figs.~\ref{fig:mSqvmsToptypeIII} and \ref{fig:msqmgII}, although shifted to larger masses.}. 
The allowed regions for the masses correspond to an
uncertainty of only $0.6$ GeV in the Higgs mass calculation. In
view of the different outputs provided by different numerical
codes (see the discussion at the beginning of Section
\ref{subsec:numsetup}), this is certainly too optimistic at
present. The allowed ranges of masses shown in Figs.~\ref{fig:mSqvmsToptypeIII} and \ref{fig:msqmgII} should therefore be
considered only as rough estimates. The red regions are for
$M_{T,\Sigma}= 10^{14}~\GeV$ and the green ones for $M_{T}=
10^{9}\,\GeV$ and $M_{\Sigma}= 10^{13}\,\GeV$. The left (right) plots
are for $A_0=0\,\TeV$ ($A_0=-5\,\TeV$). Due to the CMSSM assumptions,
stop and squark masses are tightly correlated, once the Higgs mass is
fixed. It is interesting to note that once $A_0$ is also set, the
requirement that the Higgs mass falls into the CMS window leads to
mass combinations which show a clear dependence on the seesaw
scale. Especially noteworthy is the fact that no overlap between the
regions with $M_{\Sigma}= 10^{14}~\GeV$ and $M_{\Sigma}= 10^{13}~\GeV$
exists in case of $A_0=-5~\TeV$. Similar allowed mass ranges are
obtained for seesaw type II. However, in this case we observe some
overlap between the combinations of masses, even for the extreme cases
of seesaw scales shown. It is nevertheless interesting that type III
with a scale as low as $10^{13}\,\GeV$ does not allow squark and stop
masses as large as type II does. For large values of $A_0$ and fixed
$m_{h^0}$, part of the parameter space is testable at the LHC with
$\sqrt{s}=14~\TeV$. However, the allowed combinations of squark and
stop mass for $A_0=0\,\TeV$ are completely out of range of LHC14.

Since in all our different setups large squark (and gluino) masses are
required in order to explain a $125~\GeV$ Higgs, the expectations are
that no direct signals for SUSY will be found in the near future. The
LHC reach for $\sqrt{s}=14~\TeV$ and 300 (3000) fb$^{-1}$ has been
recalculated very recently in~\cite{1207.4846}. The main conclusions
of this study are that, via gluino/squark searches, LHC14 will be able
to explore SUSY masses up to $m_{\tilde g}\sim 3.2~\TeV$ ($3.6~\TeV$)
for $m_{\tilde q}\sim m_{\tilde g}$ and of $m_{\tilde g}\sim 1.8~\TeV$
($2.3$TeV) for $m_{\tilde q}\gg m_{\tilde g}$ with 300 fb$^{-1}$ (3000
fb$^{-1}$). Thus, for $m_{h^0}\sim 125~\GeV$, only a small
part of the allowed parameter space will be probed.  However, future
plans for the LHC envisage the possibility of ramping up the
center-of-mass energy to $\sqrt{s}=33~\TeV$~\cite{Heuer:2012gi}. With
such a huge gain in energy, considerably larger regions of
the parameter space allowed in our examples would become
testable.
\begin{table}
\centering
\begin{tabular}{c c c c c c c}
\hline
 & $m_0$ [TeV] & $M_{1/2}$ [TeV] & $A_0$ [TeV] & $\tan\beta$ & sign$(\mu)$
& $M_{SS}$ [GeV] \\
\hline
Point I & 3 & 3 & 0 & 20 & + & $10^{14}$ \\
Point II & 7 & 7 & 0 & 20 & + & $10^9$ \\
\hline
\end{tabular}
\caption{Benchmark points with heavy squarks and gluino. Point I
corresponds to a type I seesaw and point II to a type II seesaw. Both
points have been chosen to give a Higgs mass of approximately
$m_{h^0}=125~\GeV$.}
\label{prosp-points}
\end{table}
\begin{table}
\centering
\begin{tabular}{c c c}
\hline
Particle & Point I & Point II \\
\hline
$\tilde{\chi}^0_1$ & 1.35 & 0.54 \\
$\tilde{d}_L$, $\tilde{s}_L$ $\tilde{u}_L$, $\tilde{c}_L$ & 6.2 & 7.3 \\
$\tilde{d}_R$, $\tilde{s}_R$ $\tilde{u}_R$, $\tilde{c}_R$ & 6.0 & 7.3 \\
$\tilde{b}_1$ & 5.6 & 6.1 \\
$\tilde{b}_2$ & 5.9 & 7.1 \\
$\tilde{t}_1$ & 4.7 & 5.0 \\
$\tilde{t}_2$ & 5.6 & 6.1 \\
$\tilde{g}$ & 6.2 & 2.7 \\
\hline
\end{tabular}
\caption{Some SUSY masses for the benchmark points given in Table
\ref{prosp-points}. All masses are given in TeV. For point I (point II)
we find $m_{h^0} \simeq 125.6$ ($125.1$)~GeV.}
\label{prosp-masses}
\end{table}
\begin{table}[t!]
\centering
\begin{tabular}{c c c}
\hline
Production cross-section & Point I & Point II \\
\hline
$\tilde{t}_1$ $\tilde{t}_1^*$ & 3.47 & 2.08 \\
$\tilde{q}$ $\tilde{q}^*$ & 8.36 & 0.60 \\
$\tilde{q}$ $\tilde{q}$ & 72.6 & 9.59 \\
$\tilde{q}$ $\tilde{g}$ & 41.0 & 793 \\
$\tilde{g}$ $\tilde{g}$ & 3.49 & 17000 \\
\hline
\end{tabular}
\caption{Most relevant production cross-sections for the
benchmark points given in Table \ref{prosp-points}. All cross-sections
are given in attobarns. These numbers have been computed with {\tt
Prospino} \cite{Beenakker:1996ed}.}
\label{prosp-xsections}
\end{table}

To check this more quantitatively, we have calculated the cross
sections for SUSY production at $\sqrt{s}=33~\TeV$ for some
representative points using the code {\tt
  Prospino}~\cite{Beenakker:1996ed}~\footnote{The calculation of SUSY
  cross section at such large c.m.s. energy requires extrapolation of
  the measured PDFs and, therefore, is probably only a rough
  estimate.}. The input parameters for two benchmark points lying
inside the CMS Higgs mass range are given in
Table~\ref{prosp-points}. We have chosen one point for type I seesaw
(point I) and one for type II seesaw (point II), although for the SUSY
production cross sections only squark and gluino masses are really
important, of course. The corresponding SUSY spectra and some
production cross sections are given in Tables~\ref{prosp-masses} and
\ref{prosp-xsections}, respectively. Point I has been deliberately
chosen to give $m_{\tilde q} \simeq m_{\tilde g} \simeq 6~\TeV$, while
point II leads to a heavier squark spectrum (around $7~\TeV$) but a
lighter gluino. From Table~\ref{prosp-xsections} one can see that
point I (point II) would yield around $\sim 40$ ($\sim 5300$)
squark/gluino events for an integrated luminosity of 300
fb$^{-1}$. These numbers are without any cuts and, therefore, should
be taken as rough estimates. Nevertheless, they serve to illustrate
how LHC33 would be able to cover most of the region of interest.  This
is also confirmed by Fig.~\ref{fig:msqmgII} where we show the allowed
regions in the $(m_{\tilde q},m_{\tilde g})$ plane with $m_{h^0}$ in
the CMS interval, and two extreme values of $M_{SS}$, for type II
seesaw (top) and type III seesaw (bottom). In the left (right) panel
$A_0=0\,\TeV$ ($A_0=-5\,\TeV$).

As before, we conclude that different seesaw models lead to distinct
allowed combinations of masses. Still, in these plots, large overlaps
between the regions for fixed $A_0$ and different $M_{SS}$ are
observed.  Nevertheless, we find it especially encouraging that in
type II and III seesaws gluino masses should be within the reach of
LHC33 in almost all cases, for a Higgs mass in the CMS preferred
window. The results also show that when $M_T=10^{9}\,\GeV$
($M_T=10^{14}\,\GeV$), $m_{\tilde g}\lesssim 4.0~\TeV$ ($m_{\tilde g}
\lesssim 6.4~\TeV$) for $A_0=0\,\TeV$.  Instead, smaller values for
the gluino mass are found if $A_0<0$. The corresponding numbers for
type III seesaw are $m_{\tilde g} \le 5.1\,\TeV$ ($m_{\tilde g} \le
5.4\,\TeV$) for $M_{\Sigma} = 10^{13}\,\GeV$ ($M_{\Sigma} =
10^{14}\,\GeV$) and $A_0=0\,\TeV$. These values should be compared
with those of type I seesaw/pure-CMSSM where gluino masses can be as
large as $m_{\tilde g} \lesssim 7\,\TeV$ for $A_0=0\,\TeV$.

A word of caution should be added to this discussion, owing to the
fact that the upper limit on $m_{\tilde g}$ shown in
Fig.~\ref{fig:msqmgII} is very sensitive to the choice of the range
for $m_{h^0}$.  In particular, if the Higgs mass is as large as
$m_{h^0}=128$ GeV, which is currently not excluded, gluino masses up
to $10\,\TeV$ and larger, would be allowed. Also, for small values of
$\tan\beta$, say in the window $\tan\beta \simeq (1-7)$, loop
corrections to the Higgs mass are known to be small. This would again
require much heavier stops and, therefore, much heavier gluinos to
explain a $m_{h^0}\simeq 125$ GeV.

\section{Concluding remarks}
\label{sec:conc}

In this work we have computed the mass of the lightest Higgs boson
within the three tree-level realizations of SUSY seesaws and studied
the main features of these models in light of the recent ATLAS and CMS
results on Higgs mass searches. We have also complemented our analysis
by considering the MEG bound on the LFV radiative decay
$\mu\to e \gamma$.  As in the pure CMSSM case, in SUSY seesaws
a Higgs mass in the range $(125-126)\,\GeV$ (as preferred currently by
CMS and ATLAS~\cite{CMStalk,ATLAStalk}) requires in all cases a rather
heavy SUSY spectrum. This is expected since $m_{h^0}$ is only
sensitive to low-energy masses and mixings, and not to high-energy
seesaw parameters (at least in a direct way). In other words, one can
in principle find a different set of input parameters for each seesaw
model leading to the same value of the Higgs mass.
For this reason, a possible seesaw discrimination cannot
rely on the Higgs mass data alone. Still, one expects
to observe some differences in the physical low-energy SUSY spectrum.

We have discussed squark, stop and gluino masses preferred by the
current Higgs data in the different seesaw scenarios. While some small
part of the parameter space allowed by a hefty Higgs will be tested at
LHC14, most of our points are beyond the reach of the next LHC
run. However, a possible increase of the LHC energy to
$\sqrt{s}=33\,\TeV$~\cite{Heuer:2012gi} would make it possible to
cover a large part of the parameter space allowed by the current Higgs
data in our models. By considering some benchmark scenarios, we have
also concluded that, in some cases, the allowed regions in the
squark/stop and squark/gluino planes do not overlap when different
values of the seesaw scale are considered or distinct seesaws are
compared.  Although this is not a general feature of the models under
study, we believe this kind of analysis may be useful in the future to
distinguish among seesaw setups and/or set limits on the input
parameters of a particular model. Complementary information coming
from the flavour sector, namely from rare decay searches, can also
play a crucial r\^{o}le in the accomplishment of this task. In
particular, upcoming data from MEG (and also from other LFV dedicated
experiments) will certainly lead to further restrictions on the seesaw
parameter space.

We would like to mention that current data \cite{CMStalk,ATLAStalk}
prefers an enhanced branching ratio for the di-photon final state;
$\sigma^{\rm obs}/\sigma^{\rm SM} =1.54 \pm 0.43$ for CMS $1.9\pm 0.5$
for ATLAS. With our heavy SUSY spectrum such an enhancement can not be
explained. However, currently this ``discrepancy'' is only of the
order of ($1-2$) $\sigma$ and thus not significant.

In this work we have not considered dark matter constraints (for a
study of neutralino dark matter in the type-II and type-III seesaw setups considered in this paper,
we address the reader to Refs.~\cite{Esteves:2009qr,Esteves:2010ff}). Although dark matter is known to provide 
powerful constraints on the SUSY parameter space, one should keep in mind that these
constraints are only valid if a standard thermal history for the
early universe is assumed (see for example \cite{Gelmini:2006pw}). 
There have also been several works devoted to the study of whether lepton 
flavour violation can be probed at the LHC (some examples within SUSY seesaw are 
\cite{Esteves:2009vg,Buras:2009sg,Abada:2011mg}). We have not 
taken this possibility into account, simply because in our framework the Higgs mass constraint leads to SUSY spectra which are too heavy to allow measuring LFV at the LHC with any reasonable statistics.

Finally, we would like to remark that, although at low-energies the
CMSSM may not seem very different from its seesaw variants, the
reconstruction of the initial conditions do drastically change from
one case to the other. In view of this, one should reflect about the
meaningfulness of fitting the CMSSM input parameters in a context
where neutrino masses cannot be explained, as it happens to be in the
MSSM. Low-energy measurements do result on different preferred regions
for the input parameters when distinct models are
considered. Obviously, this is not relevant for phenomenological
studies at low energies, but it is surely crucial for studies
addressing the dynamics behind SUSY breaking.

\acknowledgments

We thank Werner Porod and Florian Staub for helpful discussions and
assistance with SPheno. M.H. acknowledges support from the Spanish
MICINN grants FPA2011-22975, MULTIDARK CSD2009-00064 and by the
Generalitat Valenciana grant Prometeo/2009/091 and the EU~Network
grant UNILHC PITN-GA-2009-237920. F.R.J. thanks the CERN Theory
Division for hospitality during the final stage of this work and
acknowledges support from the EU Network grant UNILHC
PITN-GA-2009-237920 and from the \textit{Funda\c c\~ao para a
  Ci\^encia e a Tecnologia} (FCT, Portugal) under the projects
CERN/FP/123580/2011, PTDC/FIS/098188/2008 and CFTP-FCT UNIT
777. A.V. acknowledges support by the ANR project CPV-LFV-LHC
{NT09-508531}.


\end{document}